# Temperature-Dependent Dielectric Function of Tantalum Nitride Formed by Atomic Layer Deposition for Tunnel Barriers in Josephson Junctions

Ekta Bhatia[1], Aaron Lopez Gonzalez[2], Yoshitha Hettige[2], Tuan Vo[1], Sandra Schujman[1], Kevin Musick[1], Thomas Murray[1], Kim Kisslinger[3], Chenyu Zhou[3], Mingzhao Liu[3], Satyavolu S. Papa Rao[1], and Stefan Zollner[2]

[1]*NY Creates, Albany, NY, USA*
[2]*Department of Physics, New Mexico State University, Las Cruces, NM, USA*
[3]*Center for Functional Nanomaterials, Brookhaven National Laboratory, Upton, NY 11973, USA*

We report the dielectric functions of insulating tantalum nitride (TaN) films, deposited using atomic layer deposition (ALD) on 300 mm Si/$SiO_2$ substrates, to demonstrate their suitability as tunnel barriers in tantalum-based Josephson junctions (JJ) for superconducting quantum circuits. The temperature-dependent ellipsometric angles were measured using ALD TaN films with nominal thicknesses of 13 nm and 25 nm at an incidence angle of 70°, across photon energy ranges of 0.03 eV to 0.7 eV (80–300 K) and 0.5 eV to 6.5 eV (80–600 K). This data was used to develop a dispersion model for insulating ALD TaN films that incorporates a Tauc-Lorentz oscillator with a band gap of 1.5-1.8 eV to model the interband optical transitions. The extracted dielectric function of ALD TaN films shows an insulating behavior (mid-infrared transparency) at all temperatures and for both film thicknesses tested. ALD TaN does not exhibit infrared absorption due to free carriers, even at elevated temperatures, demonstrating its insulating nature, which is required for the tunnel barrier of the JJ in quantum applications. The results of transmission electron microscopy, including selected area electron diffraction, and X-ray diffraction are also discussed. Sputter depth-profile X-ray photoelectron spectroscopy (XPS) shows an N/Ta ratio of ~1.2 throughout the film. The lower band gap, low roughness, and thermal stability of ALD TaN compared to $AlO_x$ suggest the possibility of fabricating JJs with thicker barriers while achieving critical current densities required for qubits, better control of thickness and composition, reduced topography, and resistance to aging.

## I. INTRODUCTION

The development of quantum computing depends on fabricating reliable and efficient qubits [1,2]. A Josephson junction (JJ), incorporating a superconductor-insulator-superconductor (S-I-S) trilayer, is an essential building block of superconducting qubits [3-5]. To date, JJs have largely been fabricated using Al as the superconducting layer, along with a 1 nm thick $AlO_x$ insulator formed at room temperature by controlled exposure to oxygen. In some cases, the $Al/AlO_x/Al$ stack forming the junction is connected to Nb wiring.

Recently, Ta has gained attention as a superconducting material for qubit devices [6-8]. Coherence times improved to as high as 300 μs [6] and 500 μs [7] when the material forming the shunt capacitor was changed from Nb to Ta. In a recent report, 2D transmon qubits fabricated on a Ta-on-high-resistivity Si platform achieved lifetimes up to 1.68 ms (with millisecond-scale average lifetimes) [8]. However, these qubits used traditional $Al/AlO_x/Al$ Josephson junctions. Qubits using $Al/AlO_x$ suffer from fabrication-related process-control issues that result in varying qubit frequencies, causing significant design and process overhead to avoid cross-talk between qubits. Efforts have been made to replace room-temperature oxidation of Al with ALD $Al_2O_3$ as the tunnel barrier; however, qubit performance using such a tunnel barrier has not been established [9].

Ta-based qubits that are scalable and reliable, and have predictable performance characteristics, can be fabricated using CMOS-compatible processes in a state-of-the-art 300 mm wafer scale fabrication facility [10, 14, 16]. This requires the development of α-Ta electrodes, as shown in our earlier work on chemical mechanical planarization of α-Ta for superconducting devices [10]. It also requires the development of an insulating tunnel barrier, such as TaN, with excellent intrinsic thermal stability, resistance to aging during air-ambient storage, and a stable interface with the electrode material (α-Ta in this study). Minimal disruption of typical processes [11] in a 300 mm CMOS IC fabrication facility is another highly desirable characteristic for the tunnel barrier.

A TaN tunnel barrier also has the additional advantage that Ta sputtered onto it preferentially forms the α-Ta phase, which has a suitable superconducting transition temperature and low microwave loss characteristics, both critical for quantum applications [6, 10, 35]. ALD is a deposition technique for growing conformal films with atomic-scale control of deposited thickness, and ALD TaN is an established material in the CMOS IC industry [19]. Although several materials such as ALD $Al_2O_3$ [9], sputter-deposited AlN [12], and sputter-deposited TaN [13], have been explored as alternatives to room-temperature oxidation of Al, there have been no reports of ALD TaN as an insulating barrier in Josephson junctions apart from our recent work [14].

Spectroscopic ellipsometry (SE) is a powerful technique to quantify the dielectric response of materials [15]. In addition to SE, the ALD TaN films were characterized by atomic force microscopy (AFM) for morphology; X-ray reflectometry (XRR) and transmission electron microscopy (TEM) for thickness; and X-ray diffraction (XRD) and selected area diffraction (SAD) in the TEM to assess crystallinity. Stoichiometry and its variation through film depth were analyzed through XPS sputter depth profiling.

This paper focuses on detailed SE measurements of ALD TaN films deposited on 300 mm wafers in a wide temperature range from 80 K to 600 K. Measurements span photon energies of 0.03-0.7 eV (mid-infrared) and 0.5-6.5 eV (near-infrared, visible, and ultraviolet). These ellipsometric data, acquired at multiple angles of incidence, were used to model the dielectric function of the ALD TaN films. We also determined the across-wafer uniformity of thickness, stoichiometry, and band gap. This work demonstrates the suitability of ALD TaN deposited on $Si/SiO_2$ as a tunnel barrier for Ta- and Nb-based S-I-S JJs, with the potential to improve junction uniformity and qubit frequency stability.

## II. EXPERIMENTAL

### A. Sample preparation

A 50 nm thick silicon oxide layer was thermally grown on B-doped, double-side polished Si (100) wafers with a diameter of 300 mm and a nominal resistivity of 10 Ω.cm. An *in situ* sputter clean before TaN deposition reduced the oxide thickness by 5-15 nm (see Sec. III for details). ALD TaN films of different thicknesses (nominally 13 nm and 25 nm) were deposited on the silicon oxide layer in a 300 mm cluster tool using commercially available, proprietary chemistries. A few conditioning wafers were used before the growth of the ALD TaN film to ensure chamber stability in terms of particle count on the wafer after deposition [16]. The films were then characterized using in-line XRR, AFM, XRD, and XPS tools. The wafer was cut into coupons for offline measurements such as TEM, sputter depth profiling, and SE.

### B. Spectroscopic Ellipsometry set-up

Spectroscopic ellipsometry measurements were performed on samples from the wafer center, mid radius, and edge for two different thicknesses of ALD TaN on thermal oxide of silicon, along with control measurements of the thermal oxide and the silicon substrate to serve as inputs for modeling the ALD TaN data. A medium-sized air compressor was used to blast alumina particles onto the back surface of the samples to roughen them. Aluminum oxide of grit 220 proved to be the best choice for roughening the back surface of these samples, without causing scratches on the front surface, to reduce specular reflections of infrared wavelengths from the back surface of the silicon wafer, permitting better data acquisition without depolarization of the reflected beam and easier data modeling. The samples were then cleaned ultrasonically with distilled water and isopropyl alcohol after sandblasting. The spectroscopic ellipsometry measurements were performed on a J. A. Woollam V-VASE rotating-analyzer ellipsometer equipped with a Berek waveplate compensator and a J. A. Woollam Fourier-transform infrared (FTIR) VASE Mark II rotating-compensator ellipsometer [15,17], as shown in Fig. 1. Temperature-dependent data from 80 K to 600 K were collected inside a Janis ST-400 ultrahigh vacuum (UHV) cryostat (with quartz windows on the V-VASE and diamond windows on the IR-VASE), with liquid nitrogen for cooling and a 50 Ω Pt resistor for heating [15]. Room-temperature spectra were acquired at incidence angles from 50° to 80°, in steps of 5°, and from 0.5 to 6.5 eV with 0.02 eV steps, whereas temperature-dependent measurements were restricted to an incidence angle of 70°. Measurements were carried out with five positions of the Berek wave plate compensator and both positive and negative polarizer angles to improve accuracy. We selected a broad range of incidence angles at room temperature because the Brewster angle of $SiO_2$ (55°) is much smaller than that of Si (75°). We also acquired infrared spectra at the same angles of incidence from 0.03 to 0.70 eV with 8 $cm^{-1}$ resolution and a fixed analyzer (0° and 180°) and polarizer (±45°).

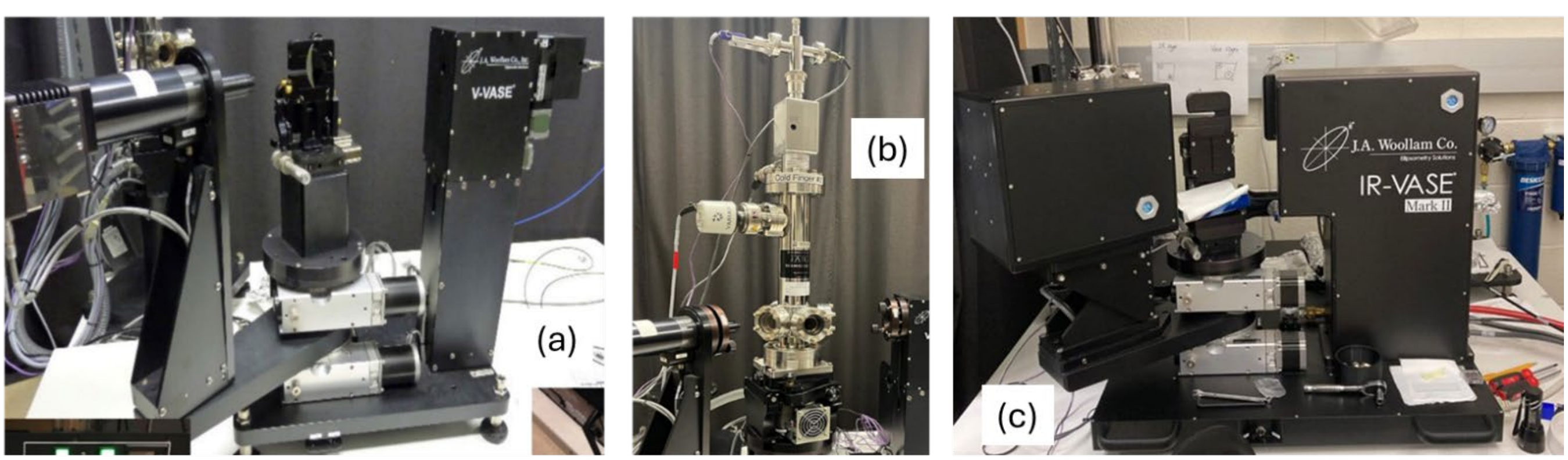

FIG. 1. Spectroscopic ellipsometry set-up: a) J. A. Woollam V-VASE ellipsometer, b) ultrahigh vacuum cryostat c) J. A. Woollam Fourier-transform infrared ellipsometer.

## C. TEM set-up

TEM analysis was performed on a coupon taken from the mid-radius region of the wafer with a 25 nm ALD TaN film. TEM lamellae were made using the *In-situ Lift-out* method with an FEI Helios G5 UX DualBeam FIB/SEM, with final Ga+ milling performed at 2 keV. These analyses were performed using a ThermoFisher Titan3 G2 80-300 system. This setup consists of an XFEG electron gun, a monochromator, and a probe spherical aberration corrector. For imaging, the microscope was configured in STEM mode, operating at an accelerating voltage of 300 kV and a beam current close to 100 pA. Imaging was performed using a 50 µm C2 aperture and a 17.9 mrad convergence angle, with simultaneous capture from both bright-field and high-angle annular dark-field (HAADF) detectors. This sample was also used for selected-area diffraction analyses conducted in the same microscope.

## D. XPS set-up

XPS measurements of the 25 nm ALD TaN film were performed on three samples (each 10 × 5 $mm^2$) taken from the wafer center, mid-radius, and edge regions. A PHI Quantera Hybrid XPS system was used for these measurements, utilizing an X-ray source with an energy of 1486.6 eV (Al Kα) and a focal spot of 200 µm. The XPS data were collected at a take-off angle of 45°. These measurements were performed at a pass energy of 26 eV and used a step increment of 0.05 eV in the selected binding-energy ranges for O1s, N1s, C1s, and Ta4f. Depth profiles were obtained by sputtering using 2 keV $Ar^+$ ions in steps totaling 30-90 s, depending on sample thickness, and measuring X-ray-generated photoelectrons from the center of the crater after each sputter step. The sputtered area was a 3 mm × 3 mm square, while the X-ray beam diameter used was 200 µm.

## III. Results and Discussion

Figure 2 shows the ellipsometric angles ψ and Δ, and the pseudodielectric function ⟨ε⟩ (data and model) for a 25 nm thick TaN sample in the center of the wafer, measured at room temperature. $tan(\Psi)$ and Δ represent the amplitude and phase of the complex Fresnel reflectance ratio for p- and s-polarized light, measured at incidence angles from 50° to 80° in 5° increments. The pseudodielectric function <ε> is calculated from the ellipsometric angles under the assumption that the sample is a bulk substrate, which ignores the layered structure. The spectral ranges in the IR from 0.03 to 0.7 eV and in the UV from 0.5 to 6.5 eV were combined into a single data set. There is no indication of free carrier absorption by the ALD TaN film, demonstrating its insulating characteristics and its suitability as a tunnel barrier in Josephson junctions for quantum applications [16]. This contrasts with the free carrier absorption observed in earlier reports on conductive ALD TaN films for CMOS IC purposes [18]. In the infrared spectral region, <ε> is dominated by molecular absorption due to Si-O vibrations, followed by a broad interference fringe between 1 and 2 eV. Since the TaN/$SiO_2$ layer stack is partially transparent, the $E_1$ peak of the Si substrate is visible in <ε> near 3.4 eV.

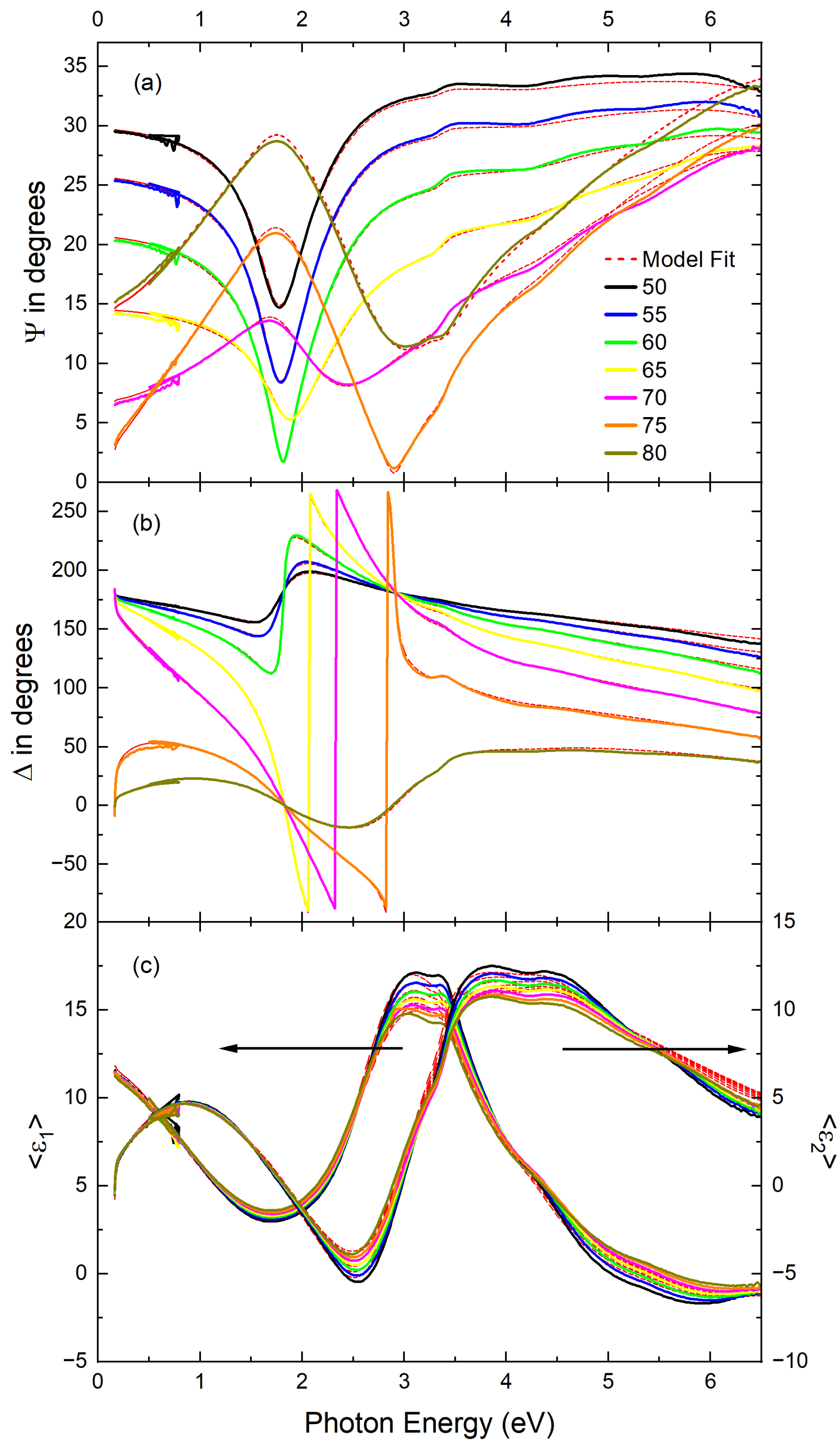

FIG. 2. Ellipsometric angles (a) Ψ and (b) Δ versus photon energy (symbols: data; lines: model) at different incidence angles, measured at room temperature for a 25 nm TaN layer on a Si substrate with a 45 nm thick thermal oxide (wafer center). Data from both ellipsometers were merged. (c) The same data are displayed as a pseudodielectric function.

In Fig. 2(c), the IR region with the $SiO_2$ lattice absorption (approximately between 0.05 and 0.2 eV) was removed from the graph because it is not relevant to this part of the research since that corresponds to the oxide layer; however, it is included in the supplementary material.

Using coupons with only thermal oxide on Si (52 nm thick as-deposited, 43 nm thick after RF sputter etching to improve adhesion), an SE dispersion model was first developed for the thermal oxide using three Gaussian oscillators at 56, 132, and 146 meV in the infrared spectral region to account for vibrations of the silicon-oxygen bonds, and a fixed pole at 11 eV with variable amplitude to model the visible and ultraviolet dispersion. This silicon-oxide model was then incorporated into the analyses of the coupons with a stack of ALD TaN on silicon oxide on bulk Si. The TaN layers were well described using a model that includes one Tauc-Lorentz oscillator with a band gap of about 1.5-1.8 eV, one or two Gaussians, and one UV pole outside our spectral range to account for transitions into excited electronic states. A Tauc-Lorentz oscillator is the most common approach to describe the optical constants of amorphous insulators such as amorphous Si, silicon nitride, etc [25]. It also works well for our TaN layers. No infrared oscillators were needed to describe the optical response of the TaN layer.

Ellipsometry can sometimes determine multiple thicknesses in a layer stack, especially if the optical constants of each layer are known. That is not the case here, since the TaN optical constants are the goal of our investigation. We use the TaN layer thickness of 25 nm determined by XRR (see Fig. 8) as an input for our ellipsometry fit. The $SiO_2$ layer thickness cannot easily be determined from XRR, since $SiO_2$ and Si have a similar scattering length density. We found that good fits to the ellipsometric angles are possible with a broad range of $SiO_2$ thicknesses ranging from 35 to 52 nm. We therefore determine the TaN optical constants at room temperature for various $SiO_2$ thickness scenarios, as shown in Fig. 3 (a). An $SiO_2$ thickness of 44 nm yields an excellent description of the region of infrared Si-O molecular vibrations near 0.15 eV, but it leads to a peak in the dielectric function of TaN near 1.2 eV, in the region of the interference fringe seen in the pseudodielectric function in Fig. 2 (c). The magnitude of this peak increases if we assume an $SiO_2$ thickness of 52 nm. It nearly disappears for a $SiO_2$ thickness of 35 nm. The magnitude of the main TaN peak around 4 eV depends much less on the $SiO_2$ thickness. The strong dependence of the peak at 1.2 eV on the $SiO_2$ thickness in our fit is suspicious. Most likely, this peak is an artifact that stems from an incomplete removal of interference effects. Since this peak disappears if we assume an $SiO_2$ thickness of 35 nm, a $SiO_2$ thickness of 35 nm is the most likely scenario and we fix it in our subsequent analysis work. The dielectric function of TaN at room temperature resulting from the fit (with an $SiO_2$ thickness of 35 nm) is shown in Fig. 3 (a). We note that the "ellipsometry thickness" (35 nm) of the oxide layer between TaN and the Si substrate disagrees with the "TEM thickness" for the same layer (43 nm) or the ellipsometry thickness of the oxide layer without TaN (also 43 nm). In our experience, such a discrepancy is not uncommon and may be related to the different decay and dispersion mechanisms of an electron wave and a light wave across an interface [32, 33].

To study the temperature dependence of the dielectric function of TaN, we performed infrared ellipsometry measurements (from 0.03 to 0.7 eV) at 80 K, 190 K, and 300 K as well as visible/UV ellipsometry measurements (from 0.5 eV to 6.5 eV) at 80 K, 190 K, 300 K, 400 K, 500 K, and 600 K at a single angle of incidence (70°) in a UHV cryostat. The angle of incidence is limited to 70° by the orientation of the cryostat entrance and exit windows. Because of reflection losses at these windows and other experimental reasons, the spectral range in the UV was more limited for measurements inside the cryostat than in air. For the analysis of the resulting ellipsometric angles, the temperature dependence of the dielectric function of bulk Si was taken into account based on the data of Franta et al. [26, 27].

As shown in Fig. 3(b), the temperature dependence of the dielectric function of TaN is moderate. The band gap of TaN (more precisely, the Tauc gap of TaN extracted from our Tauc-Lorentz oscillator model) decreases from 1.8 eV at 80 K to 1.5 eV at 600 K. A similar decrease of the band gap with increasing temperature is seen for other materials [15]. Furthermore, as expected, the broadening of the main peak around 3.5 eV increases with increasing temperature [15], associated with a moderate decrease of the high-frequency dielectric constant at 0.5 eV. The imaginary part of ε remains small at all temperatures, indicative of the insulating properties of the TaN layer.

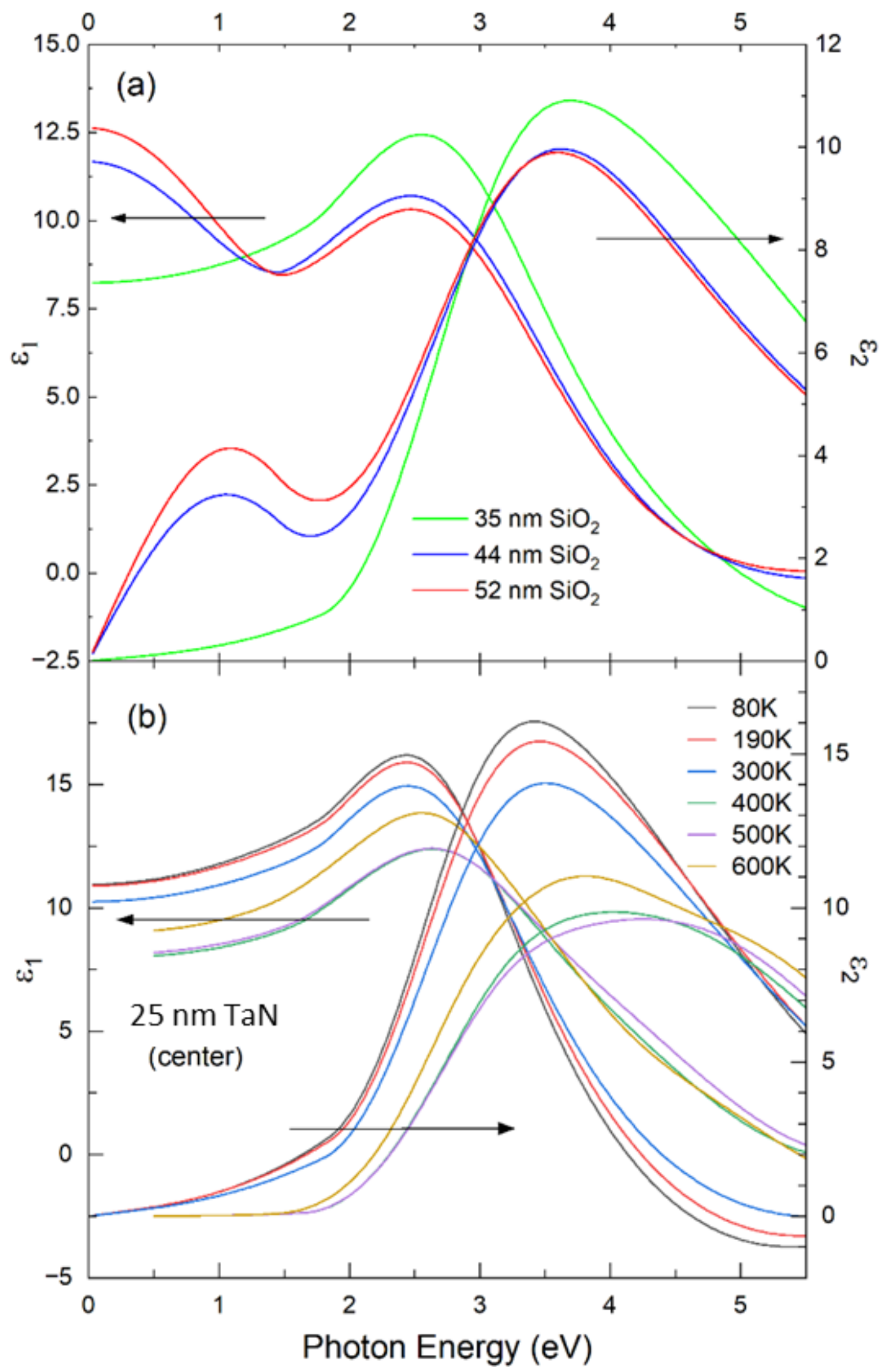

FIG. 3. (a) Real (left) and imaginary (right) parts of the dielectric function of a 25 nm thick TaN layer evaluated for different assumptions of the $SiO_2$ thickness: 52 nm (red), 44 nm (blue), and 35 nm (green). (b) Real and imaginary part of the dielectric function of a 25 nm ALD TaN film versus photon energy using data acquired at 70˚ at different temperatures from 80 K to 600 K, assuming an $SiO_2$ thickness of 35 nm.

Figure 4 shows the dielectric functions of TaN layers with thicknesses of 25 nm and 13 nm. The dielectric functions for both thicknesses follow the same overall trend as a function of photon energy, which is expected since they are both the same material. The imaginary part of the dielectric function for both thicknesses indicates a band gap near 1.7 eV at room temperature. A reduction of the dielectric function below the band gap (high-frequency dielectric constant) in thinner layers was also observed in ZnO layers grown by ALD and in $SrTiO_3$ layers grown by molecular beam epitaxy, but not in sputtered $SnO_2$ layers [34]. In general, the high-frequency dielectric constant is proportional to the electron density and inversely proportional to the band gap. It might also be affected by oxidation of the TaN. All three possibilities are ruled out by band gap measurements with ellipsometry (see previous paragraph), combined with the XRR and XPS results described below. Another possible explanation for the decrease in the high-frequency dielectric function is screening of the excitonic (Sommerfeld) enhancement of the optical constants due to built-in electric fields arising from interface charges, or because the TaN thickness becomes comparable to the excitonic Bohr radius in TaN. We did not investigate this in more detail, but we note that the optical constants of TaN are likely thickness- and process-dependent. The value of band gap is similar at the wafer center, mid-radius, and edge for both thicknesses, indicating uniform optical constants across the wafer.

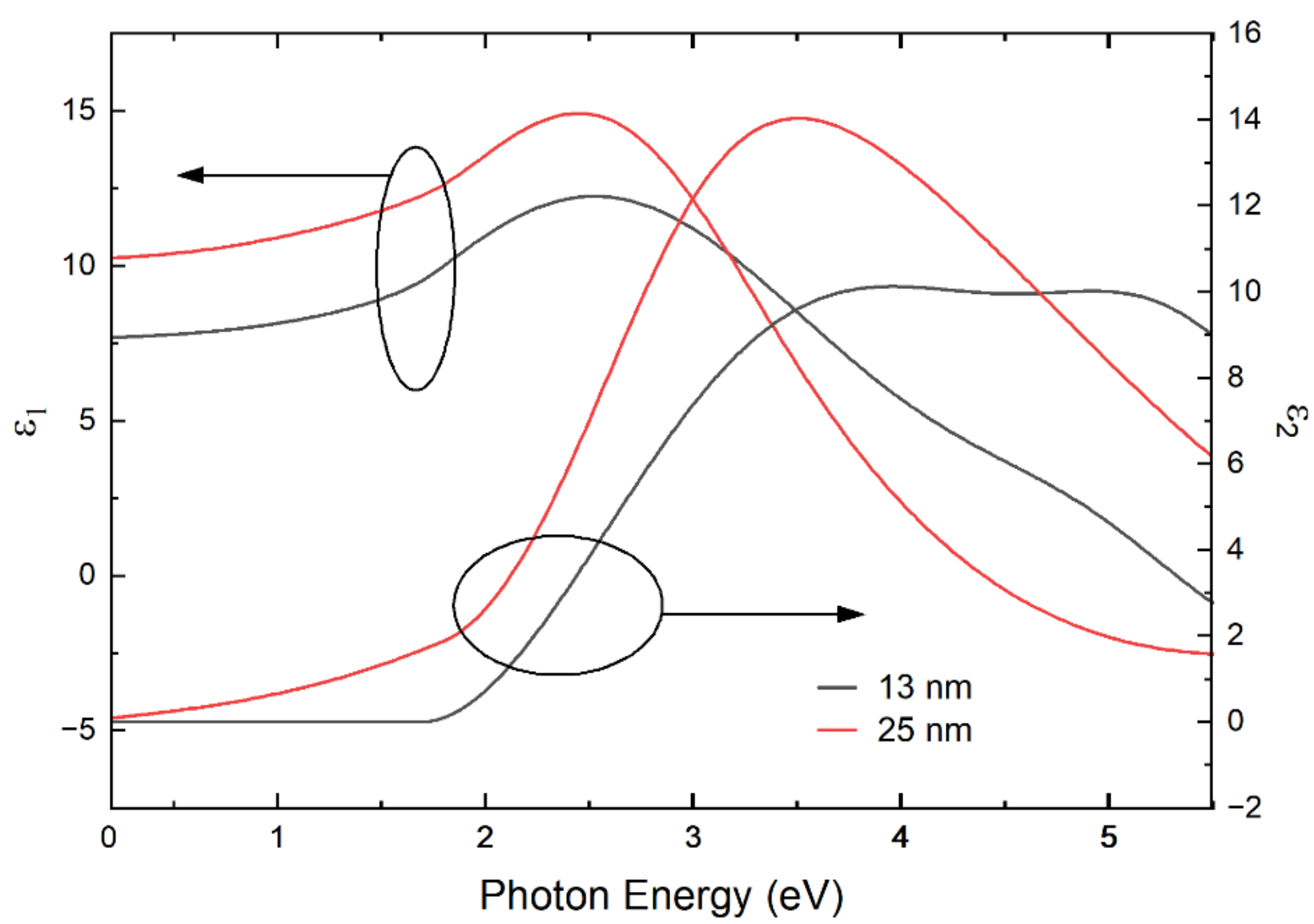

FIG. 4. Real and imaginary parts of the dielectric function of the 13 nm and 25 nm thick TaN layers, assuming an $SiO_2$ thickness of 35 nm.

Figure 5(a) shows a nearly linear relationship between the number of ALD cycles and the thickness of ALD TaN for thicknesses (determined from TEM) ranging from 1 to 25 nm. Figure 5(b) presents a TEM image of the ALD TaN layer, confirming its thickness as 25 nm. The $SiO_2$/TaN interface in Fig. 7(b) is clean, with no signature of an interfacial layer.

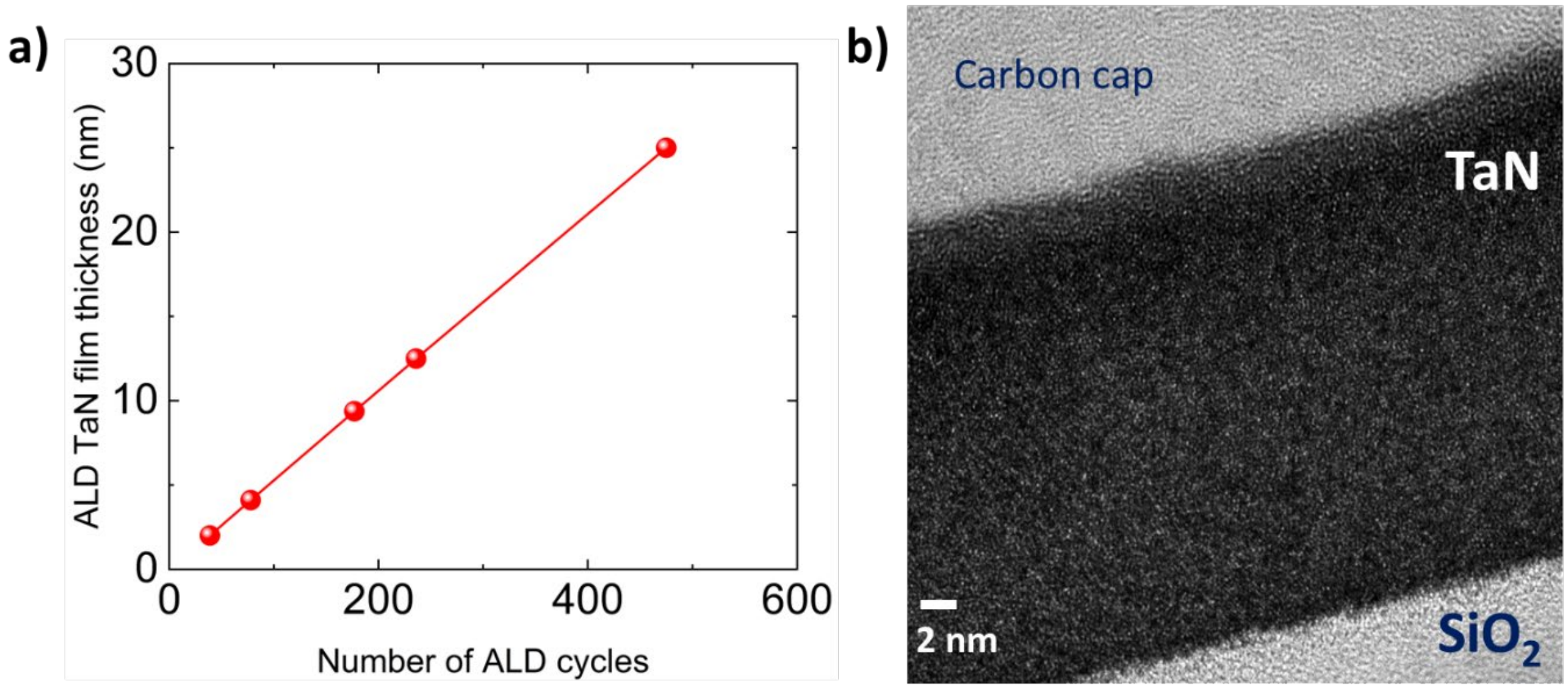

FIG. 5. a) ALD TaN thickness versus number of ALD cycles, b) TEM image showing the ALD TaN film and $SiO_2$/TaN interface.

XRR modeling was done using a stack of Si/thermal oxide/ALD TaN and a proprietary genetic algorithm that allowed the density and thickness of the different films to vary within specified limits to optimize the fit. Figure 6 shows the results of the XRR analysis for thickness determination at different locations on the wafer spanning from the wafer center to the wafer edge for two different thicknesses of ALD TaN. The within-wafer thickness nonuniformity (standard deviation as a percentage of the median) is less than 1% for 25 nm ALD TaN and ~1.1% for 13 nm ALD TaN. The variation in the extracted XRR thickness as a function of (X, Y) coordinates on the wafer is shown in Fig. 6(a). Figures 6(b) and 6(c) show the XRR raw data and fit for the 13 nm and 25 nm ALD TaN samples at the wafer mid radius location (0 mm, 70 mm). The thickness results from XRR modeling are consistent with the values obtained from the TEM cross-sections and from ellipsometry.

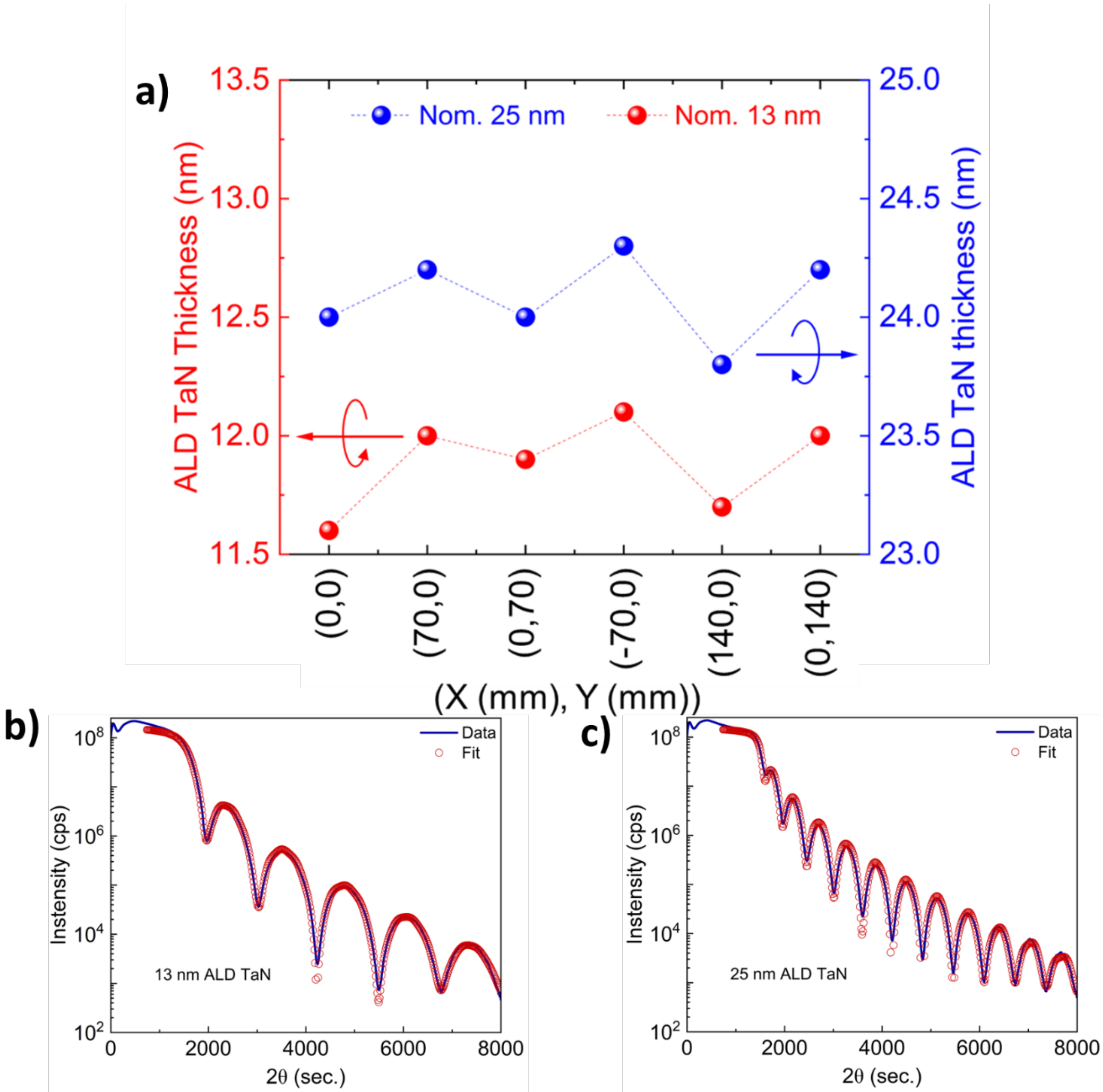

FIG. 6. Thickness determination from XRR across the 300 mm wafer: a) thickness as a function of location (X, Y) on the wafer for both nominal thicknesses, b) Example fit for XRR data for 13 nm ALD TaN wafer at (0 mm, 70 mm), c) example fit for XRR data for 25 nm ALD TaN wafer at (0 mm, 70 mm).

Figure 7 shows AFM traces of the 25 nm ALD TaN film at different locations across the 300 mm wafer. The roughness values are <0.5 nm at all points on the wafer. This is consistent with the roughness values extracted from the XRR measurements. This roughness (as well as the thickness of the native oxide on TaN) is small and can be neglected in the ellipsometry analysis.

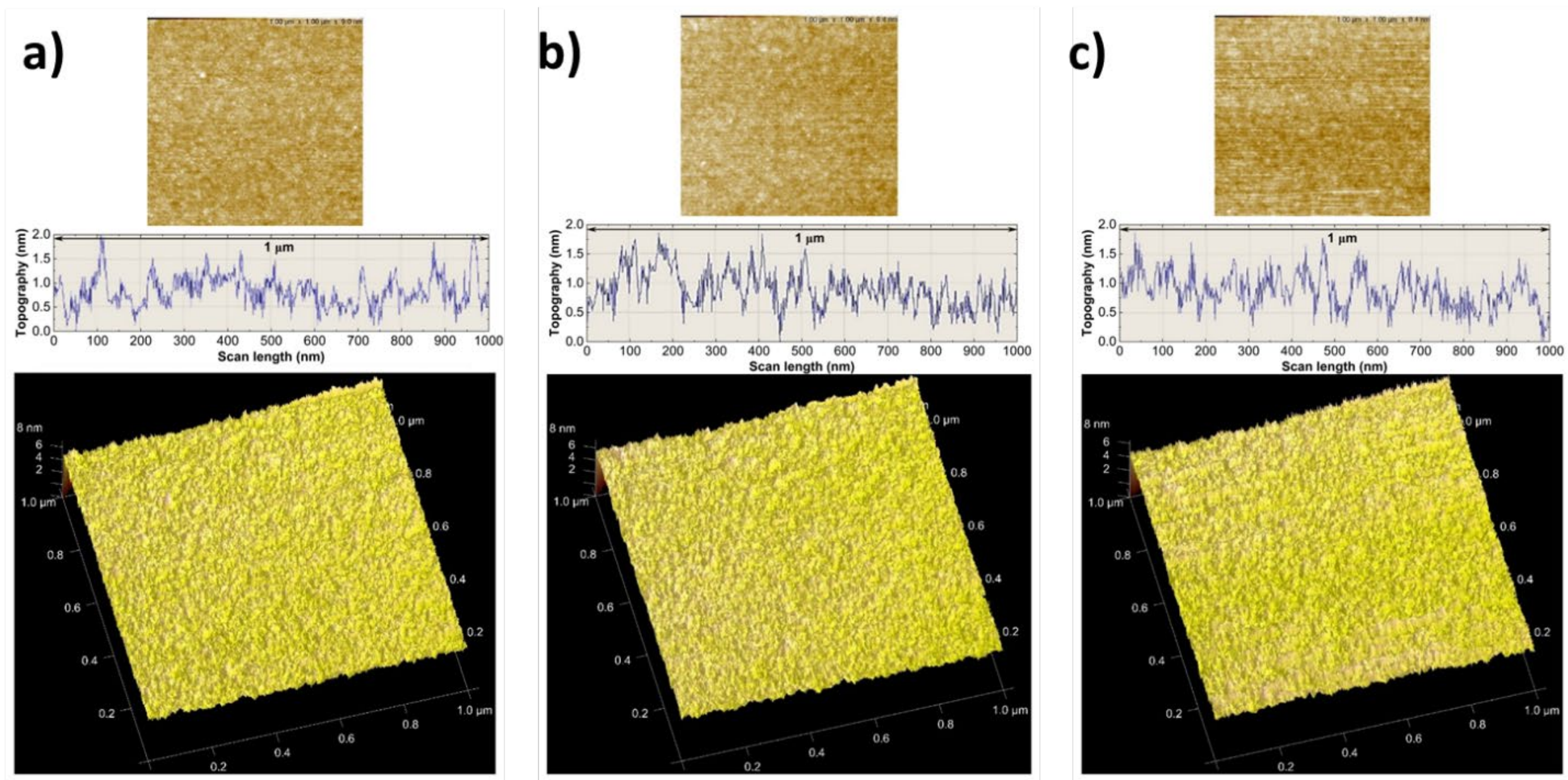

FIG. 7. AFM scans of the 25 nm ALD TaN film over a 1 × 1 µm$^2$ area at (a) wafer center, (b) wafer mid radius (0 mm, 75 mm), and (c) wafer edge (0 mm, 140 mm). In each case, the top picture shows the 2D scan, the middle picture shows the line scan (topography versus scan length), and the bottom picture shows the 3D AFM scan.

Figure 8 shows selected-area diffraction (SAD) patterns acquired using a 300 keV TEM beam from the ALD TaN film and the Si substrate. The diffraction spots in the SAD pattern of TaN (Fig. 8(a)) show the crystalline structure of $Ta_5N_6$ [30]. The twofold symmetry observed in the SAD pattern is consistent with the hexagonal structure observed along the (110) zone axis. This symmetry was confirmed at multiple locations on samples taken from the center and edge of the wafer; two examples are shown in Fig. 8(a,b). From these SAD patterns, we determined the value of the lattice constant *a* to be 5.25 Å for ALD TaN. In Fig. 8(b), we confirmed that the SAD pattern for the Si substrate is attributable to the (110) zone.

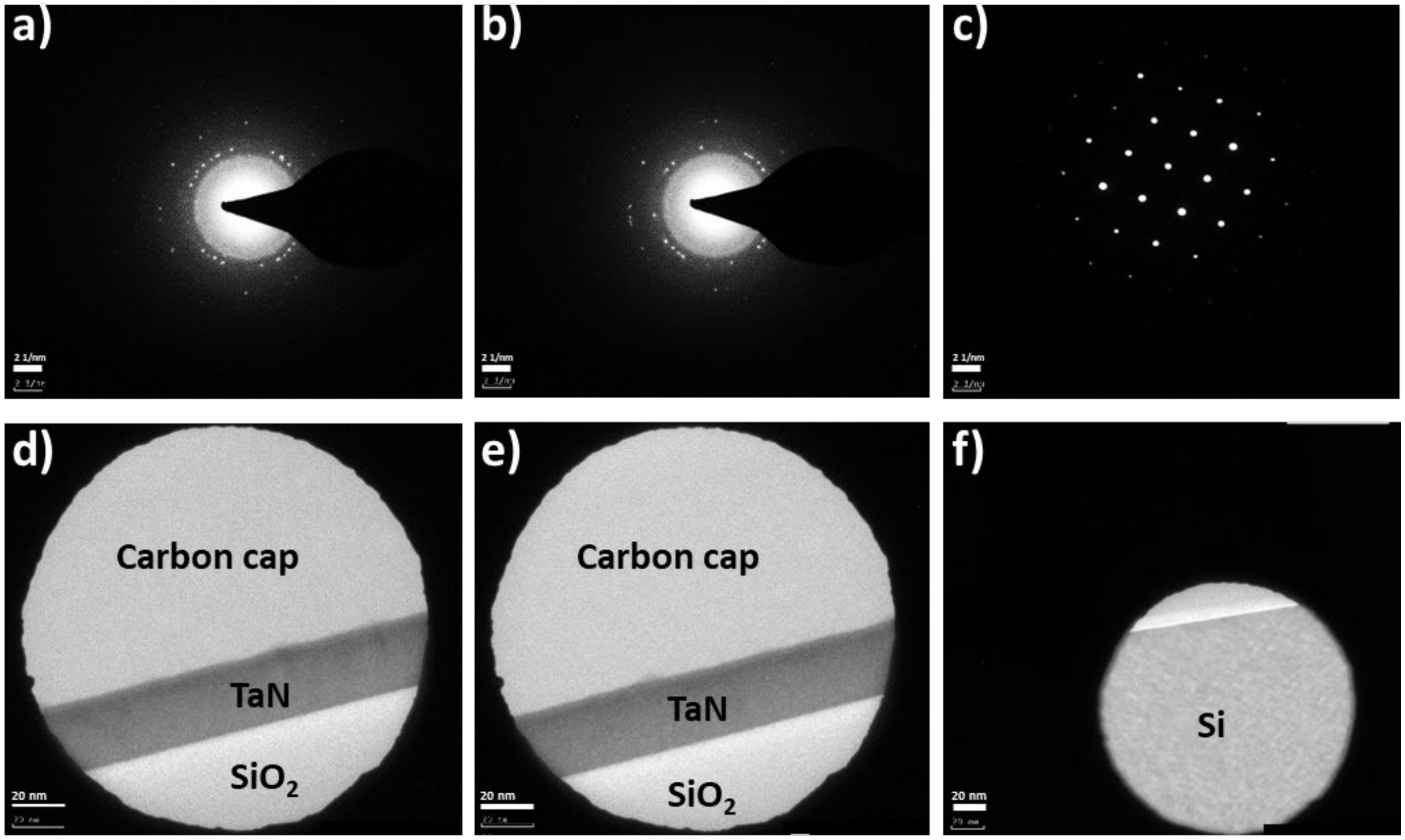

FIG. 8. Selected area diffraction (SAD) patterns using TEM from different layers. SAD patterns from (a) only ALD TaN from wafer center, (b) ALD TaN from a different location on a TEM lamella taken from the wafer center, and (c) the Si substrate. Corresponding TEM images showing the selected area aperture sizes are shown in (d)-(f).

If a thin layer (such as our TaN) is mostly amorphous, but contains a few crystalline regions, then only the crystalline regions will be seen in XRD or SAD. On the other hand, the amorphous regions will dominate the ellipsometry spectra, which can be modeled using a Tauc-Lorentz oscillator model for amorphous materials. Therefore, the presence of diffraction spots in our TaN layer is not necessarily in conflict with the use of an optical dispersion model for an amorphous material. A fully crystalline (or poly-crystalline) material is likely to show several sharp features in its optical constants.

Figure 9 shows the in-plane and out-of-plane XRD data of Si/$SiO_2$/ALD TaN and Si/$SiO_2$. The XRD spectra were obtained, in a glancing incidence configuration using Cu Ka radiation in a commercially standard 300 mm wafer-scale metrology tool. After subtracting the peaks associated with Si/$SiO_2$, there is a broad peak associated with ALD TaN with a 2θ value of 33.86˚, corresponding to a lattice constant value of 5.27 Å for *a* (indexed to the hexagonal (110) reflection). This value is consistent with the JCPDS database for hexagonal TaN with card number 00-153-9256. However, it is not possible to determine the *c* value of lattice constant based on just one broad XRD peak, and this difficulty with making a definitive assignment of a specific phase has also been seen in previous literature for $Si/SiO_2$/ALD TaN [19]. XRD analysis supports a hexagonal $TaN_x$ structure (as determined by SAD and XPS N/Ta ratio) for the ALD TaN films deposited on $Si/SiO_2$.

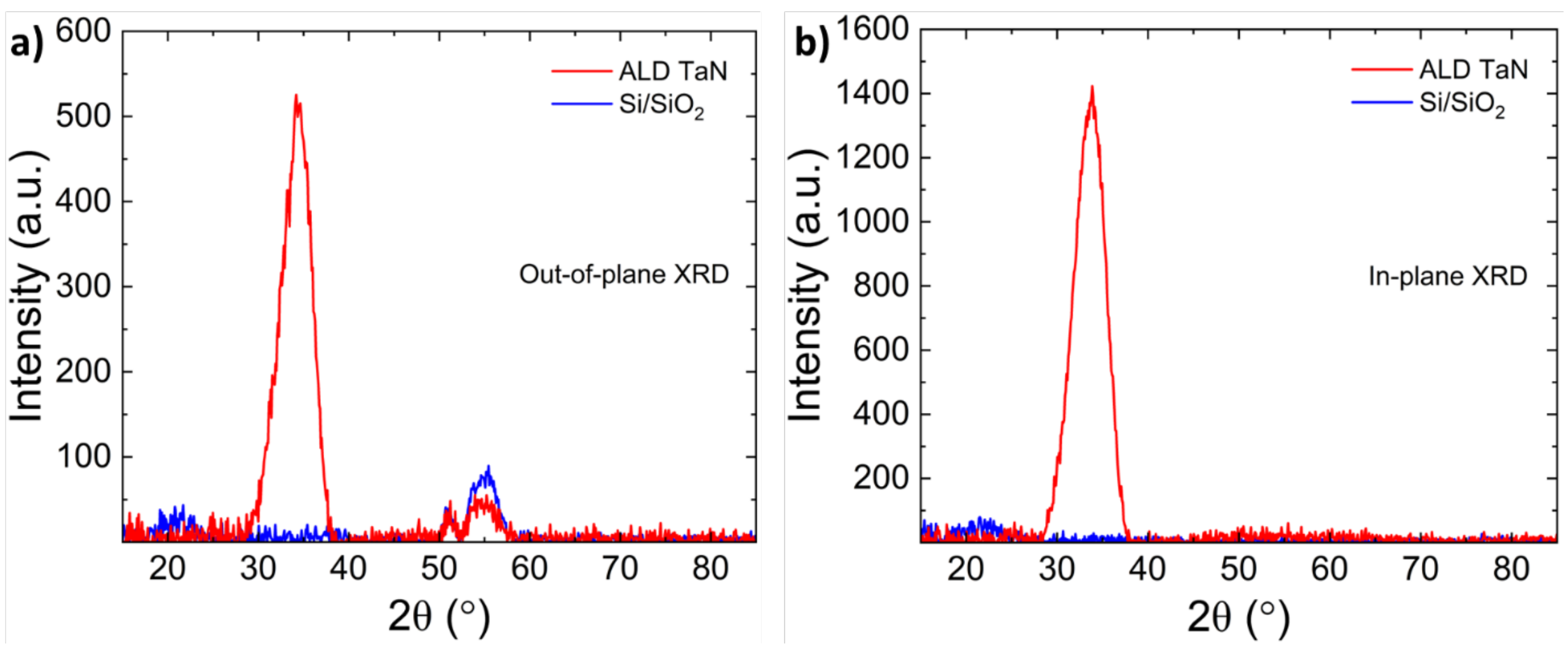

FIG. 9. XRD of 25 nm ALD TaN versus $Si/SiO_2$: a) out-of -plane XRD, b) in-plane XRD.

Figure 10 shows TEM images with zoomed-in view of the ALD TaN layer and the $Si/SiO_2$ interface. Figure 10(a) shows a high-magnification image of the TaN film. The $Si/SiO_2$ interface in Fig. 10(b) is clean and shows the crystalline lattice of Si.

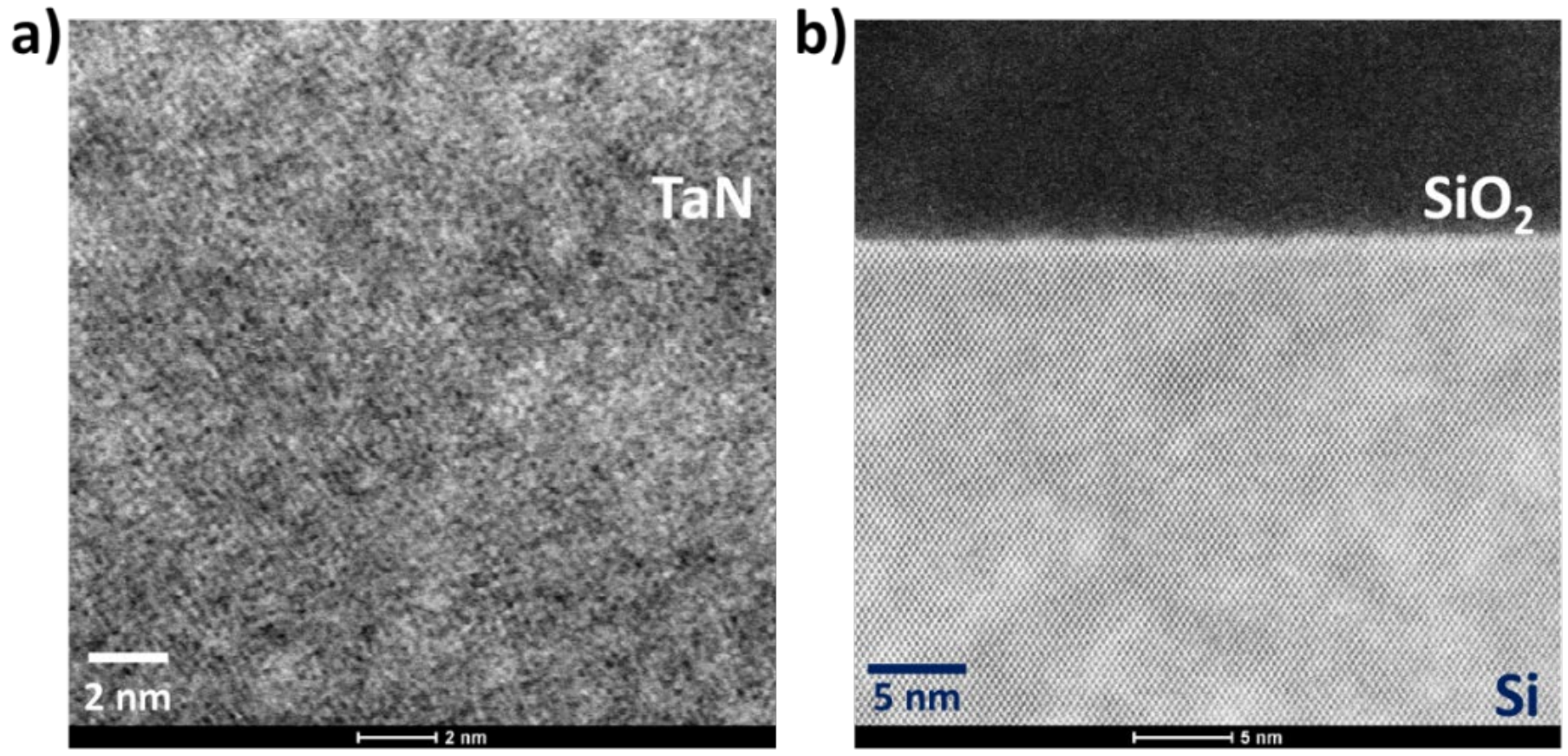

FIG. 10. TEM images showing the zoomed-in views of a) ALD TaN, b) the Si/$SiO_2$ interface.

Figure 11 shows sputter-XPS analysis using CasaXPS software for 25 nm ALD TaN films measured on coupons from the wafer center at different sputter times (corresponding to different depths from the surface of TaN). We focus on the Ta4f, O1s, C1s, Ta4p, and N1s regions for XPS analysis. A Shirley-type background was used for background subtraction in CasaXPS. Peaks were fitted using Gaussian/Lorentzian product line shapes [20-22]. The Ta4p and N1s peaks were fitted following XPS guidelines from the literature, and inelastic mean free path (IMFP) calculations were used to correct the Ta4p and N1s peak areas to determine the N/Ta ratio [21,22]. The N/Ta ratio is ~1.2 at all depths within the TaN, determined from the ratio of the corrected total area of the Ta4p peak to the corrected total area of the N1s peak after fitting. It is important to note that no detectable amounts of carbon and oxygen were seen by XPS deeper in the ALD TaN film. The N1s peak position (395.99 eV) and Ta4p peak position (403.38 eV) are consistent with the literature [23, 31]. XPS spectra for coupons taken from different locations on the wafer overlap with each other, indicating good across-wafer uniformity in TaN composition.

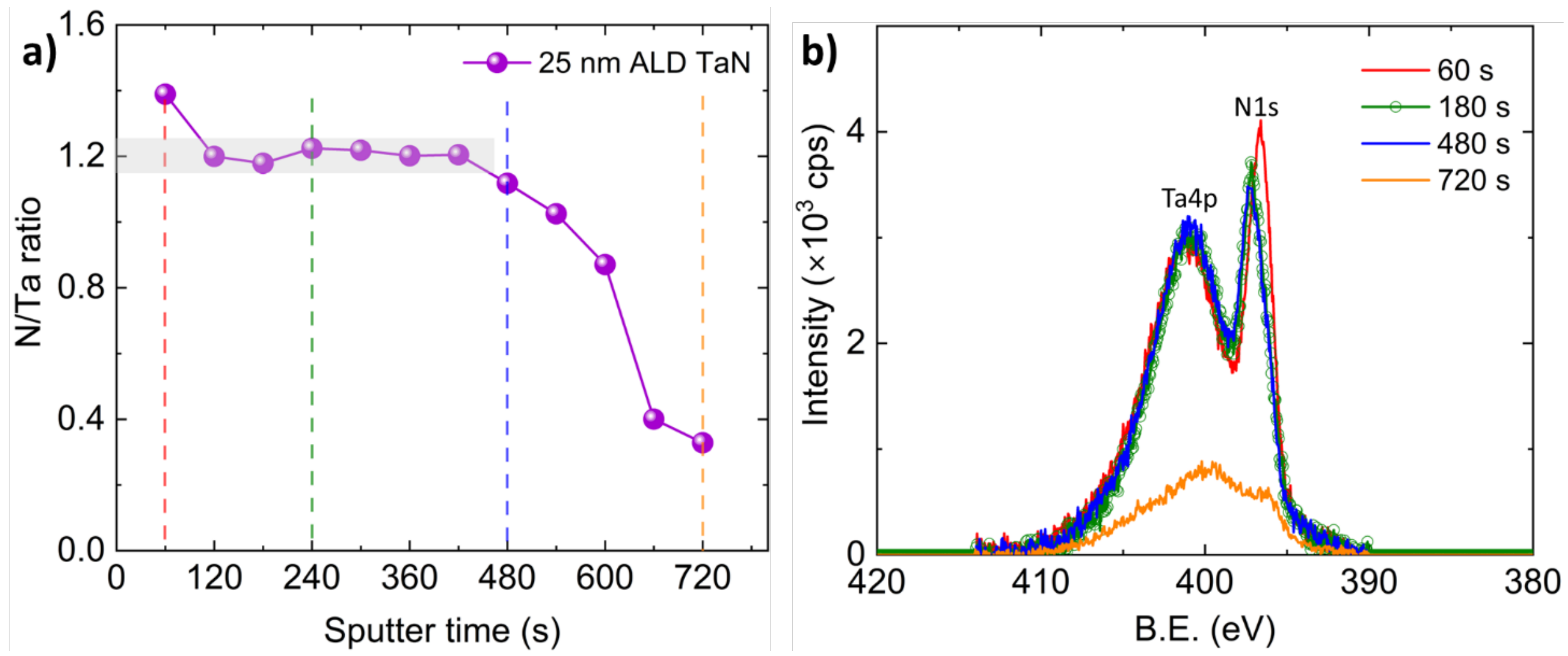

FIG. 11. Extracted N/Ta ratios as a function of sputter time from sputter-XPS, b) XPS spectra at different sputter times shown by dashed lines in (a).

## CONCLUSION

In this study, we deposited and characterized ALD TaN films and evaluated their suitability as tunnel barriers in Josephson junctions for superconducting quantum circuits. XRR, TEM, XPS, SE, and AFM measurements show excellent across-wafer uniformity in thickness, composition, roughness, and optical properties of ALD TaN films across 300 mm wafers. The bandgap of the ALD TaN is 1.5-1.8 eV, with no evidence of free-carrier absorption in the infrared, consistent with the insulating nature of the ALD TaN film. XRD and SAD analyses indicated a $Ta_5N_6$ hexagonal structure. Sputter-XPS shows an N/Ta ratio of ~1.2 and no detectable carbon or oxygen in the film. These characteristics demonstrate the suitability of ALD TaN for integration into Josephson junctions for quantum computing and superconducting logic circuits.

The ALD TaN band gap was determined to be 1.5-1.8 eV, showing only a small change in the imaginary part of the dielectric function around 1 eV when exposed to different temperatures. This demonstrates its advantages over traditional aluminum oxide barriers of higher band gap used in tunnel barriers. The smaller barrier height allows for thicker films to be utilized for the same magnitude of Josephson tunneling current, improving process control margin. The excellent surface roughness and band gap uniformity across the 300 mm wafer illustrate ALD TaN as a potential material for Josephson junctions to achieve thermally stable, reliable, and repeatable superconducting qubit devices.

Future work will focus on quantifying dielectric loss in ALD TaN. Its integration as a tunnel barrier in Josephson junctions has already been demonstrated in our recent work [14]. Next steps include evaluating coherence times in superconducting qubit devices fabricated using these Josephson junctions.

## ACKNOWLEDGMENTS

This work was supported by the Louis Stokes Alliance for Minority Participation (LSAMP) NSF-HRD #1826758. This work was also supported by the Air Force Office of Scientific Research (AFOSR) under Award No. FA9550-20-1-0135 and FA9550-24-1-0061. The support of this study by the Air Force Research Lab (AFRL), Rome, NY, through Contract Nos. FA8750-19-1-0031 and FA864921P0773, is gratefully acknowledged. We are grateful to Daniel Franta and Thomas E. Tiwald for their assistance with describing the temperature dependence of the dielectric function of bulk Si from 80 to 600 K. This research used Electron Microscopy facility of the Center for Functional Nanomaterials, which is a U.S. Department of Energy, Office of Science User Facility, at Brookhaven National Laboratory under Contract DE-SC0012704. This work was also supported by the U.S. Department of Energy, Office of Science, National Quantum Information

Science Research Centers, Co-design Center for Quantum Advantage (C2QA), under Contract No. DE-SC0012704, including Subcontract No. 390040.

## Data Availability Statement

The data that support the findings of this study are available from the corresponding author upon reasonable request.

# Supplementary Material:
# Temperature-Dependent Dielectric Function of Tantalum Nitride Formed by Atomic Layer Deposition for Tunnel Barriers in Josephson Junctions

Ekta Bhatia[1], Aaron Lopez Gonzalez[2], Yoshitha Hettige[2], Tuan Vo[1], Sandra Schujman[1], Kevin Musick[1], Thomas Murray[1], Kim Kisslinger[3], Chenyu Zhou[3], Mingzhao Liu[3], Satyavolu S. Papa Rao[1], and Stefan Zollner[2]

[1]*NY Creates, Albany, NY, USA*
[2]*Department of Physics, New Mexico State University, Las Cruces, NM, USA*
[3]*Center for Functional Nanomaterials, Brookhaven National Laboratory, Upton, NY 11973, USA*

The supplementary material shows raw ellipsometry data, which were summarized and discussed in the main article.

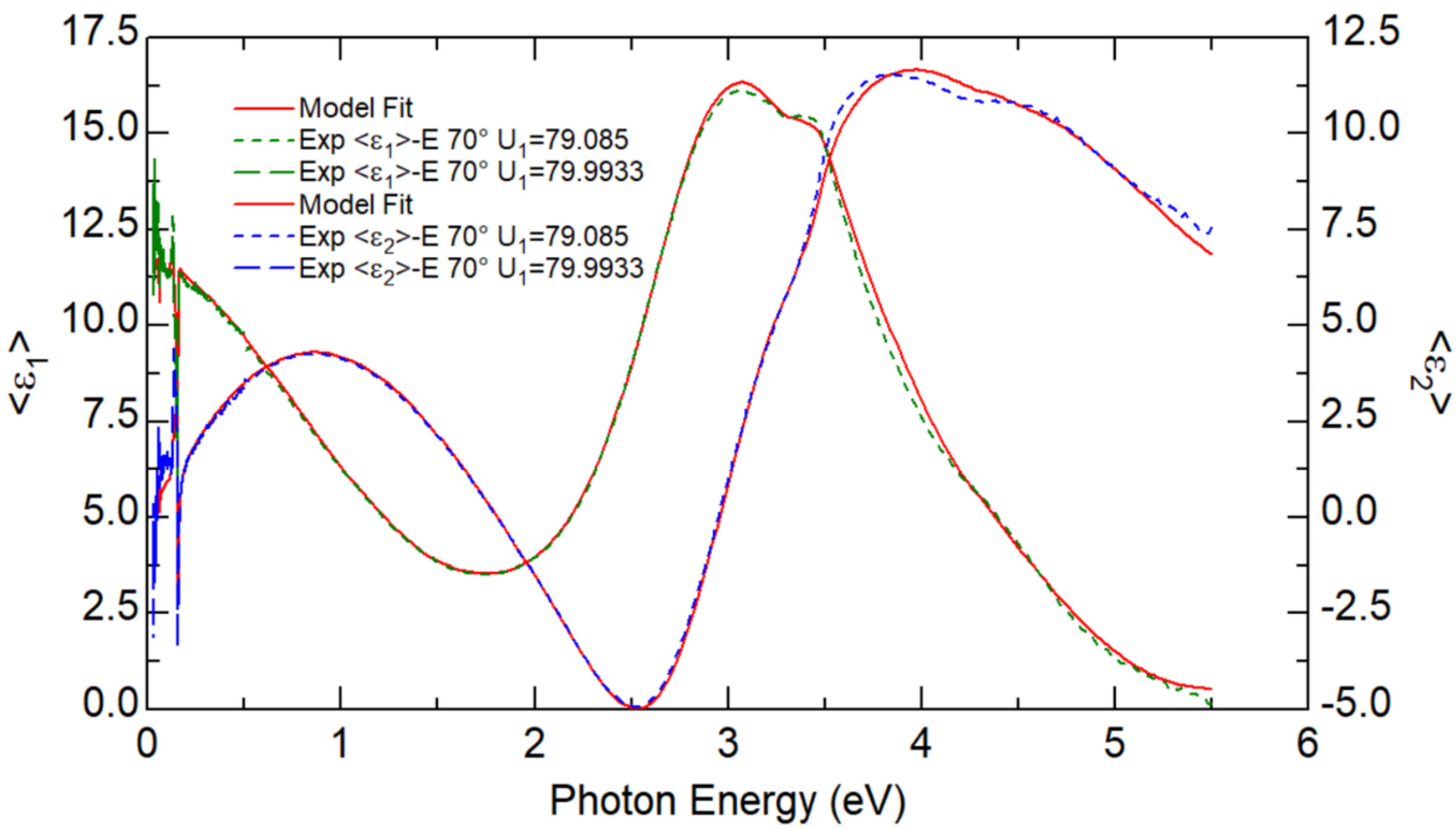

FIG. S1: Real and imaginary part of the pseudodielectric function for 25 nm of TaN and 35 nm of $SiO_2$ at 80 K in the wafer center (symbols: data; lines: fit).

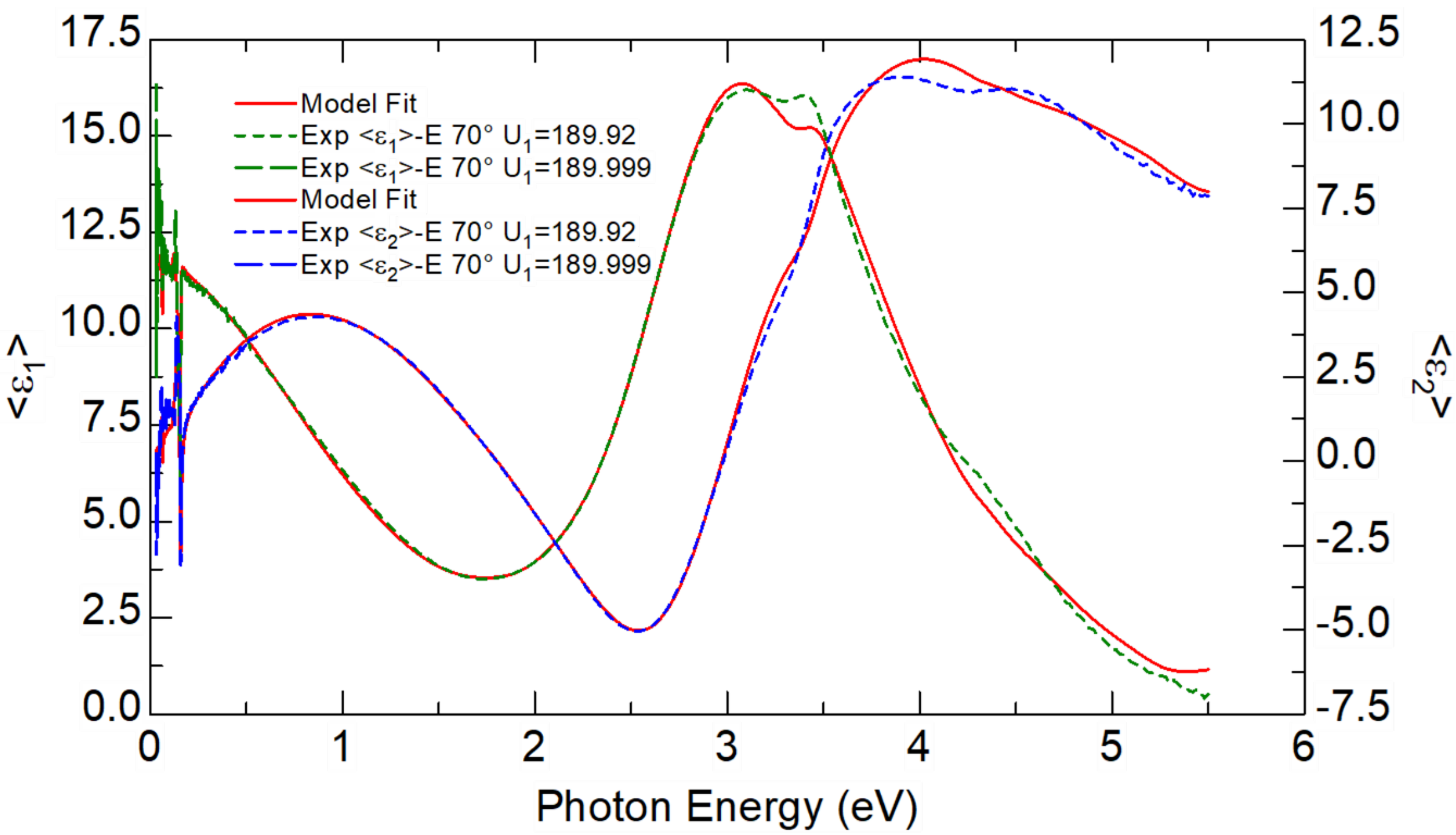

FIG. S2: Real and imaginary part of the pseudodielectric function for 25 nm of TaN and 35 nm of $SiO_2$ at 190 K in the wafer center (symbols: data; lines: fit).

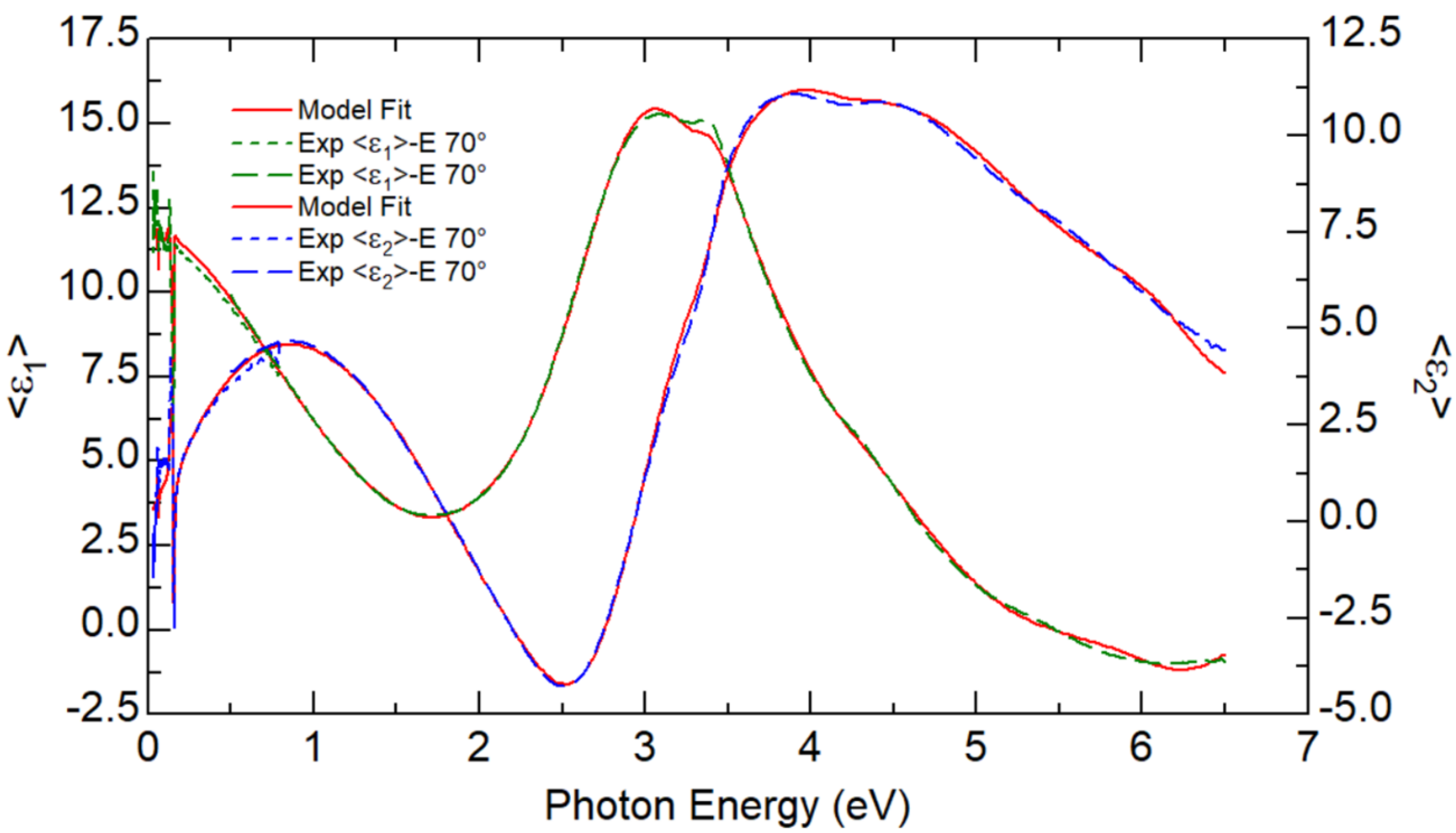

FIG. S3: Real and imaginary part of the pseudodielectric function for 25 nm of TaN and 35 nm of $SiO_2$ at 300 K in the wafer center (symbols: data; lines: fit).

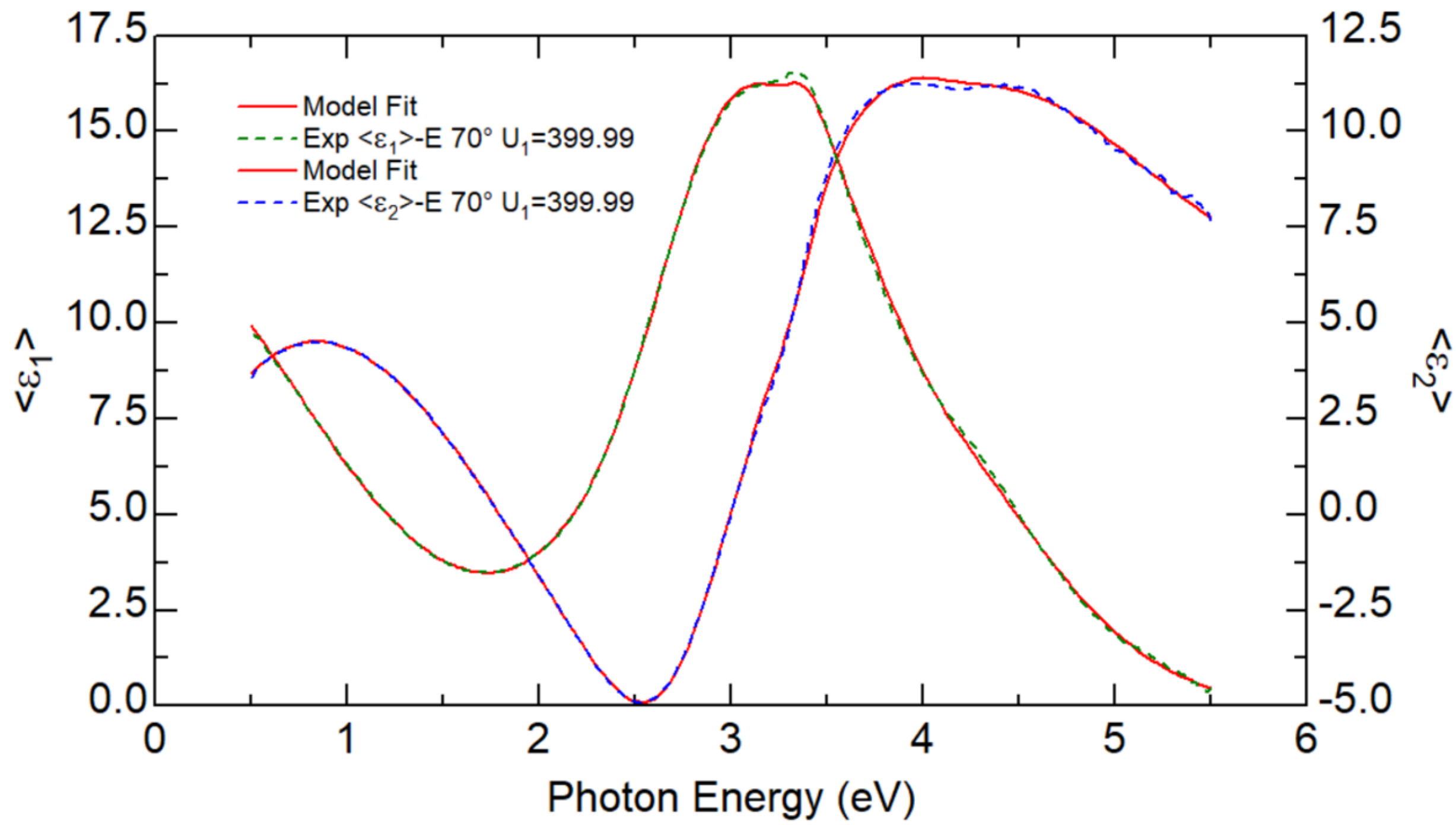

FIG. S4: Real and imaginary part of the pseudodielectric function for 25 nm of TaN and 35 nm of $SiO_2$ at 400 K in the wafer center (symbols: data; lines: fit).

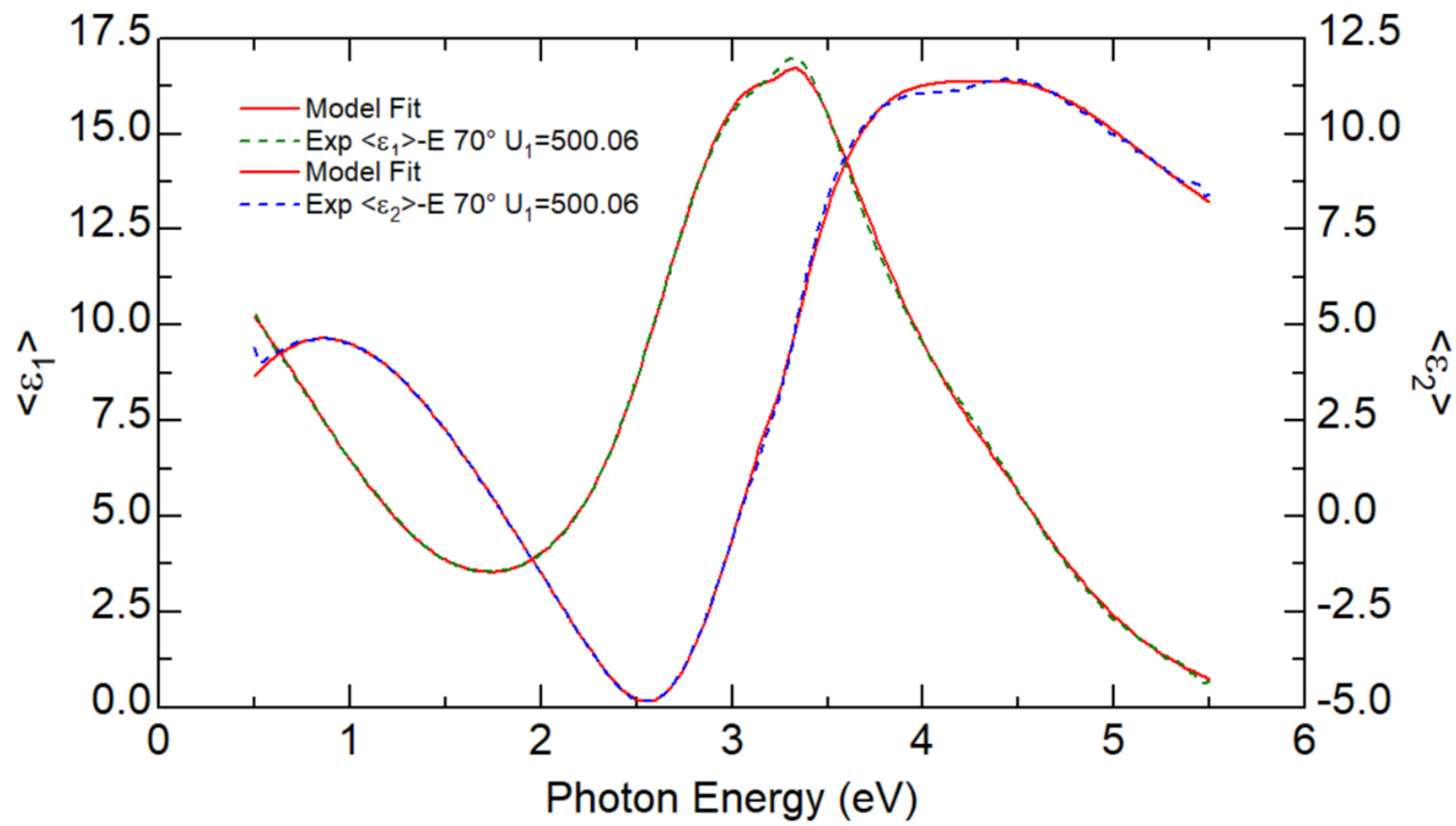

FIG. S5: Real and imaginary part of the pseudodielectric function for 25 nm of TaN and 35 nm of $SiO_2$ at 500 K in the wafer center (symbols: data; lines: fit).

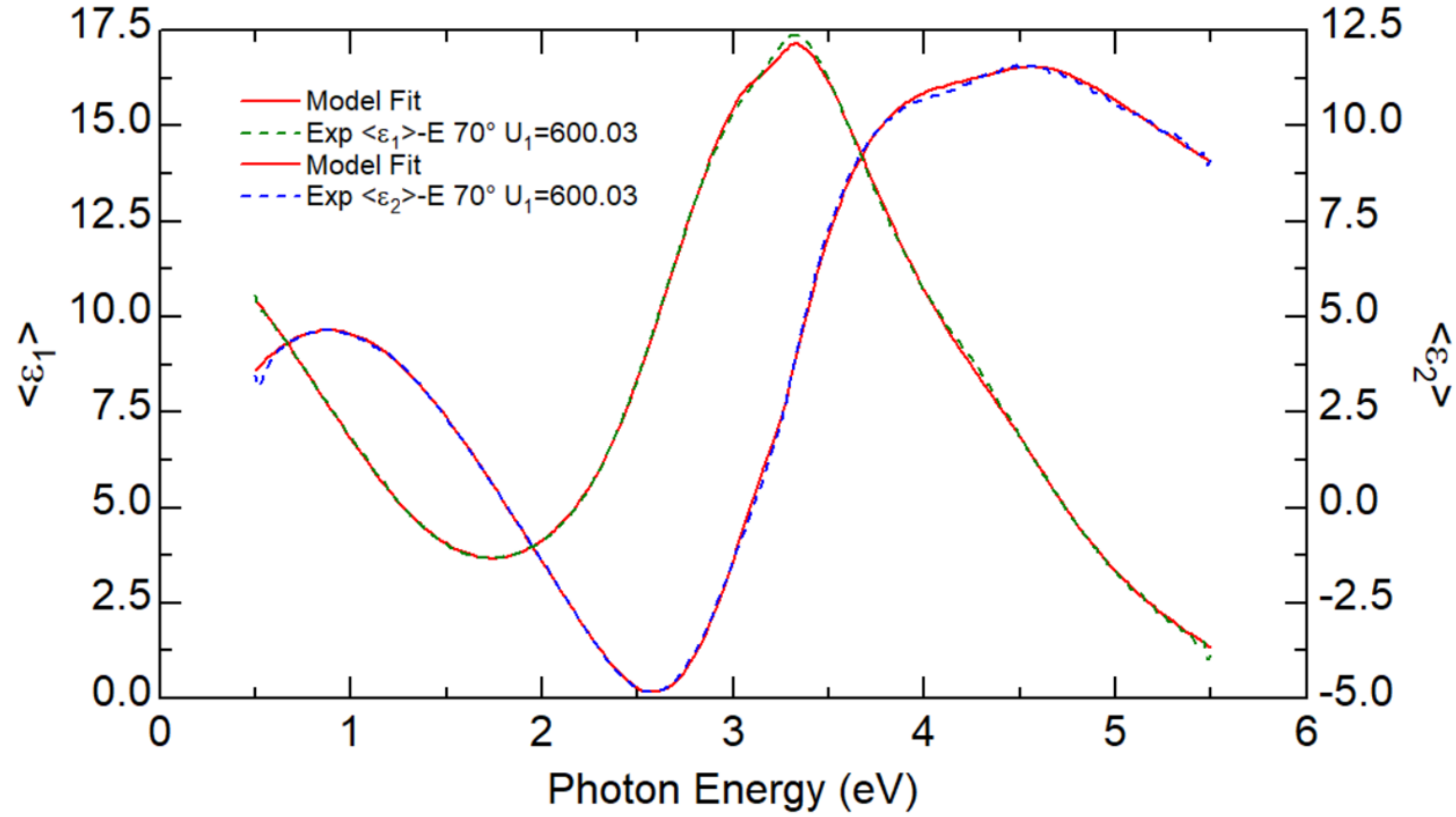

FIG. S6: Real and imaginary part of the pseudodielectric function for 25 nm of TaN and 35 nm of $SiO_2$ at 600 K in the wafer center (symbols: data; lines: fit).

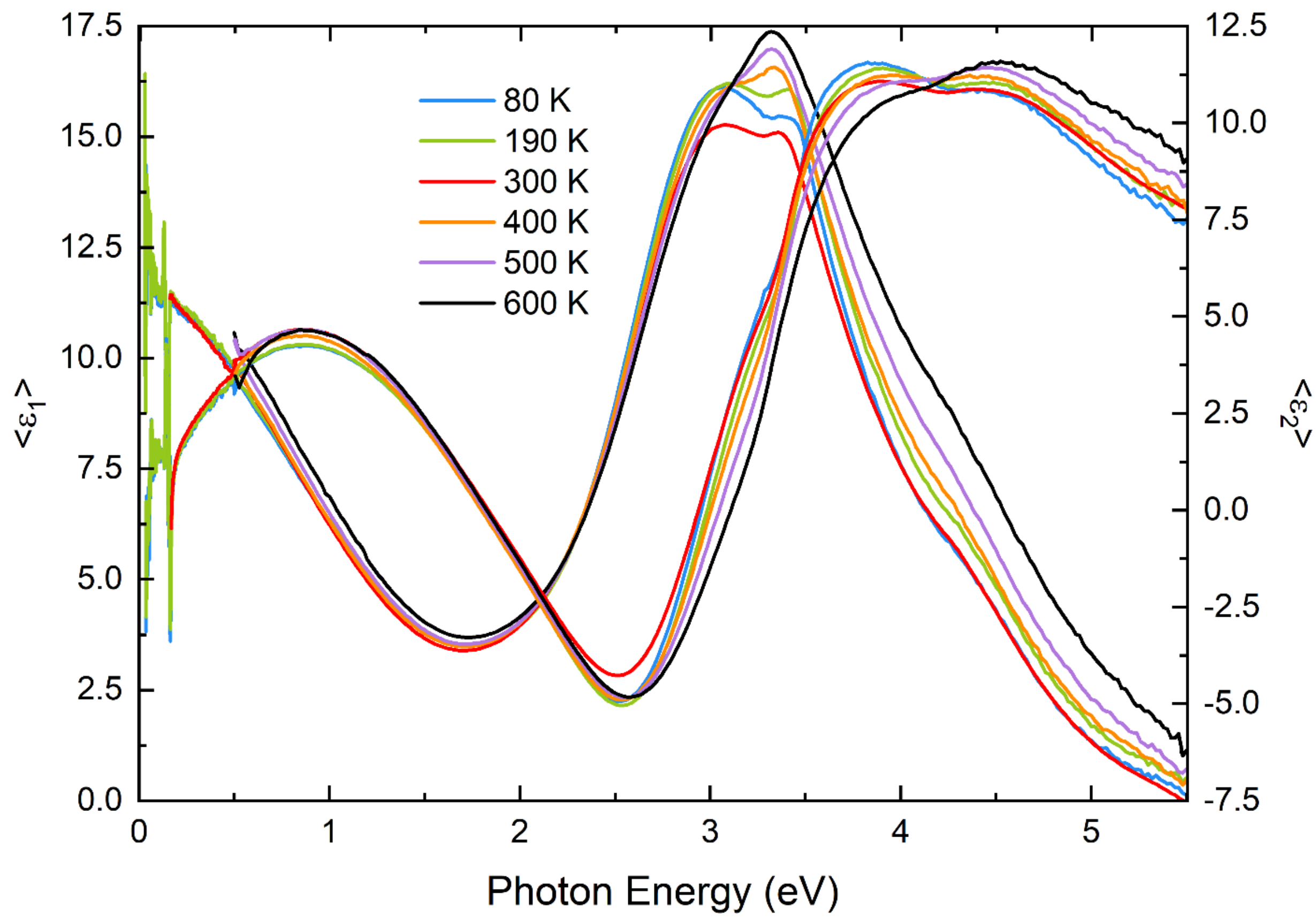

FIG. S7: Real and imaginary parts of the pseudodielectric function for 25 nm of TaN and 35 nm of $SiO_2$ from 80 K to 600 K in the wafer center (only experimental data are shown).

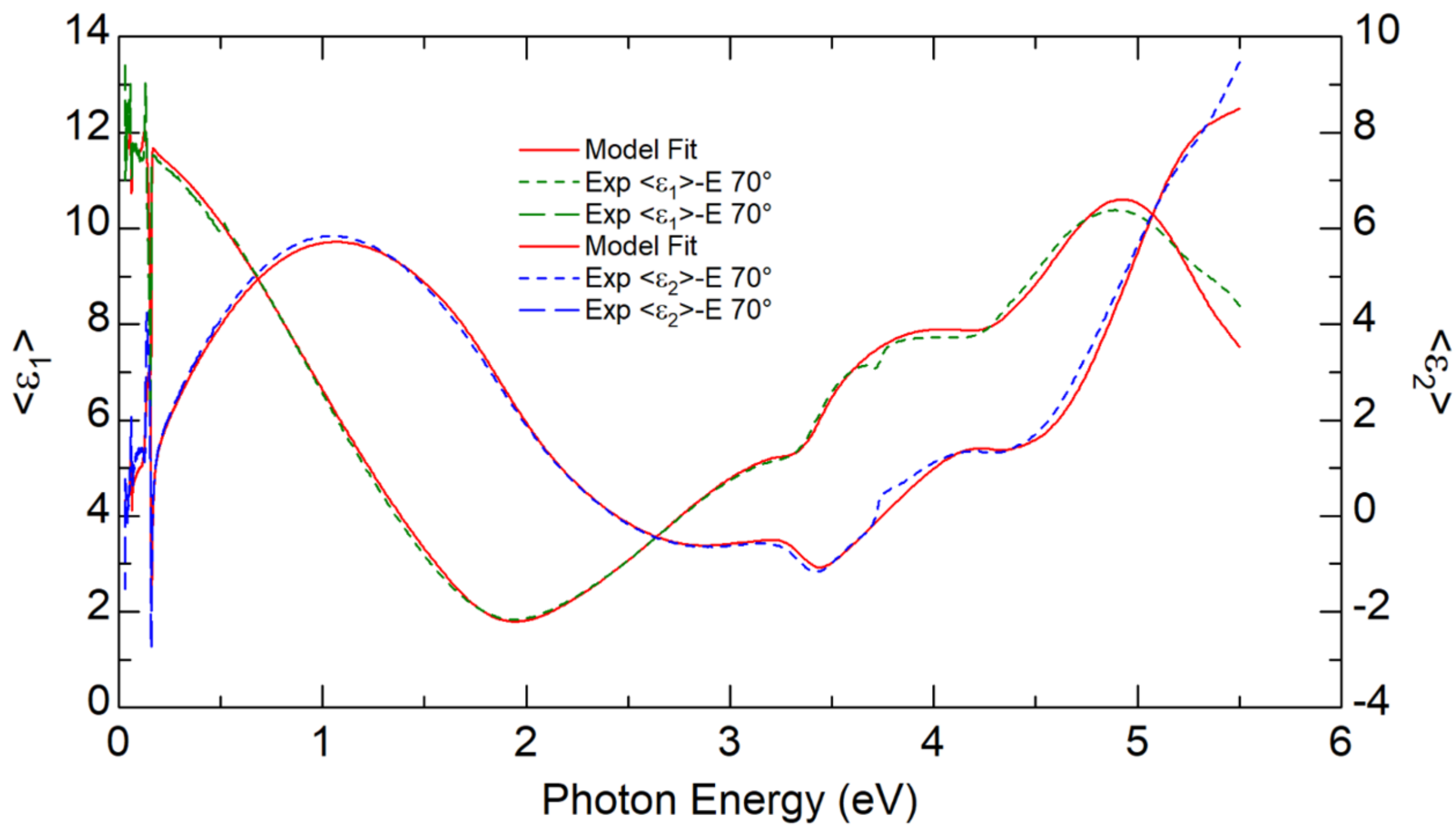

FIG. S8: Real and imaginary part of the pseudodielectric function for 13 nm of TaN and 35 nm of $SiO_2$ at 300 K in the wafer center (symbols: data; lines: fit).

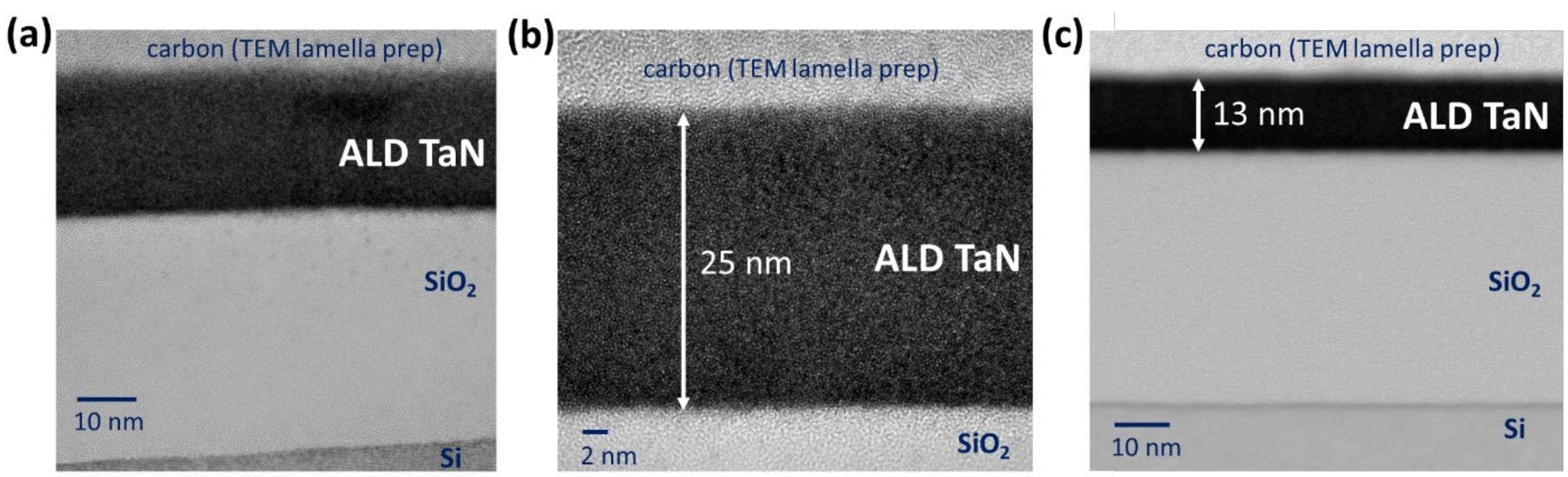

FIG. S9: TEM cross-sections confirming the film thickness at wafer center for film stacks with 25 nm and 13 nm ALD TaN.

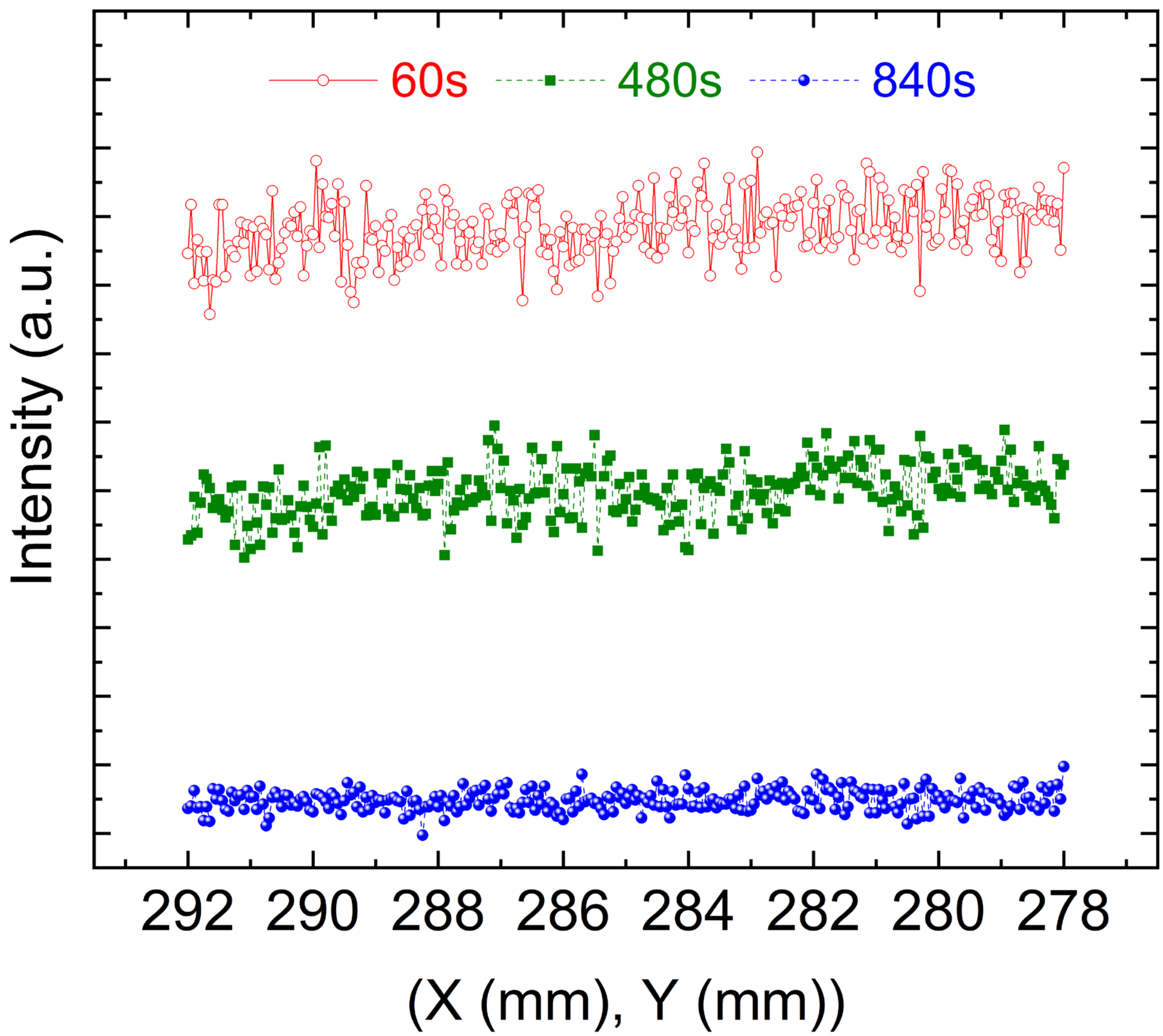

FIG. S10: SputterXPS data at different sputter times and at different depths inside the TaN film showing no carbon in the ALD TaN.

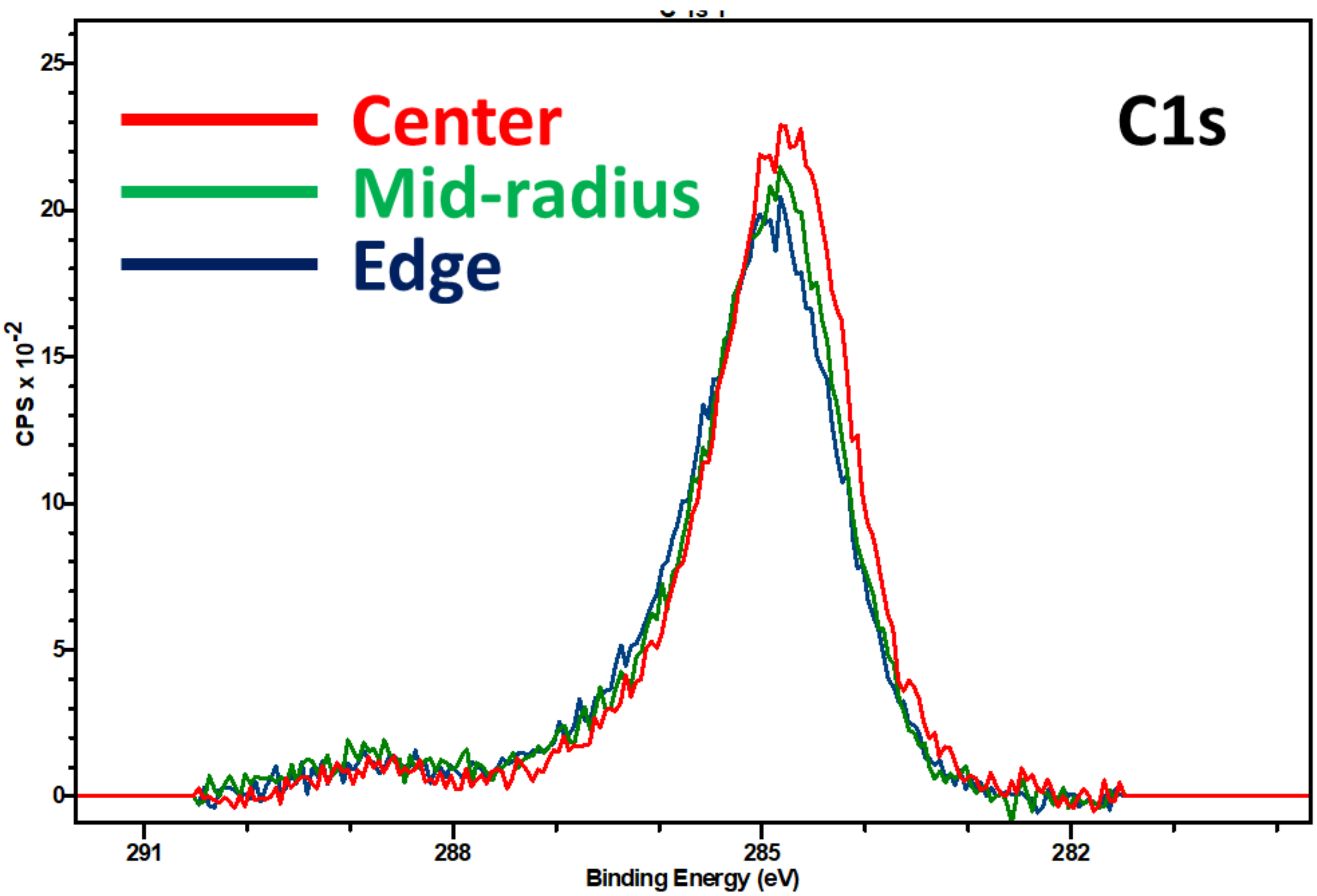

FIG. S11: XPS data (background subtracted) showing carbon on the surface at all the locations of the wafer.

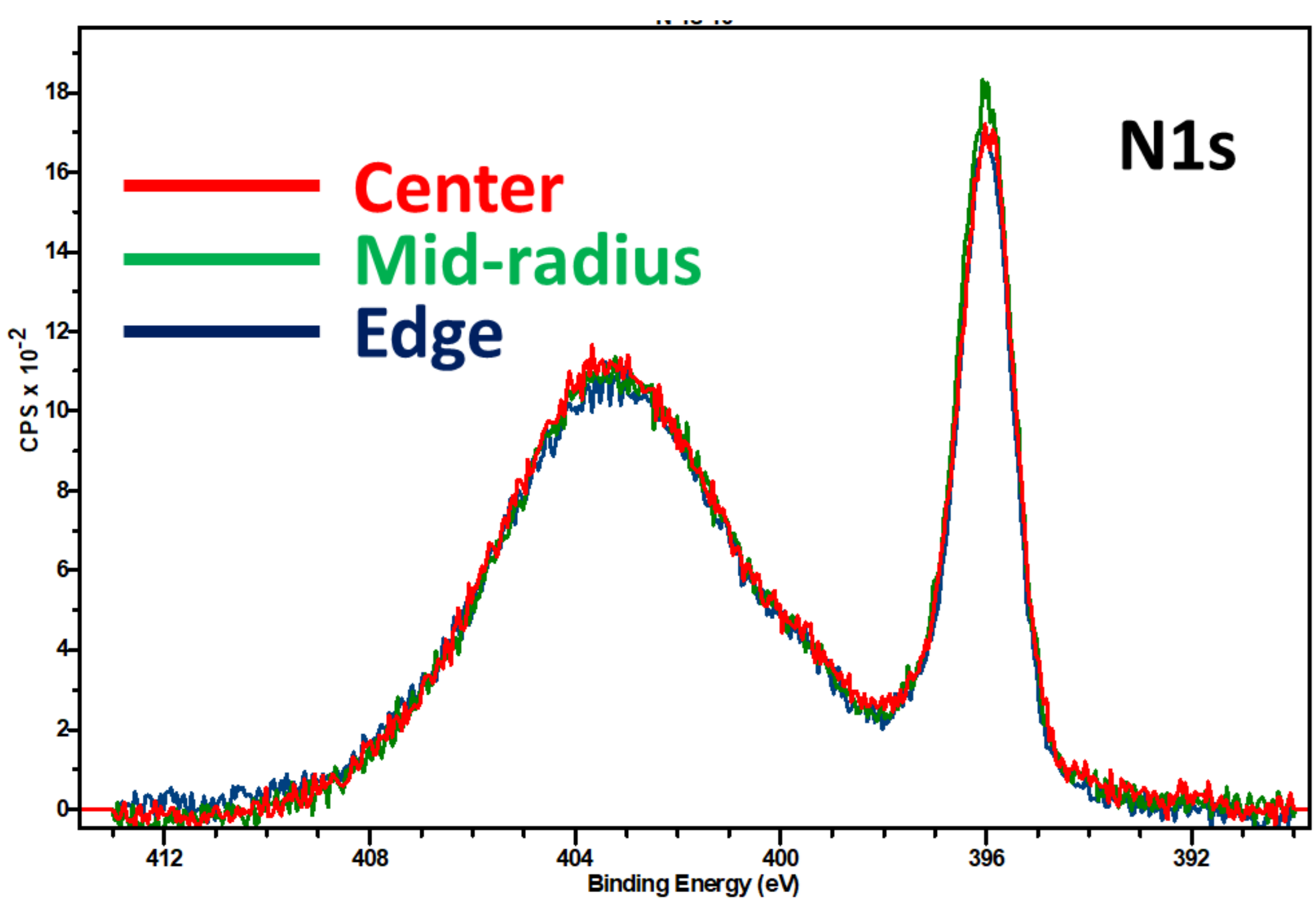

FIG. S12: XPS data (background subtracted) taken from surface showing uniformity across the wafer.

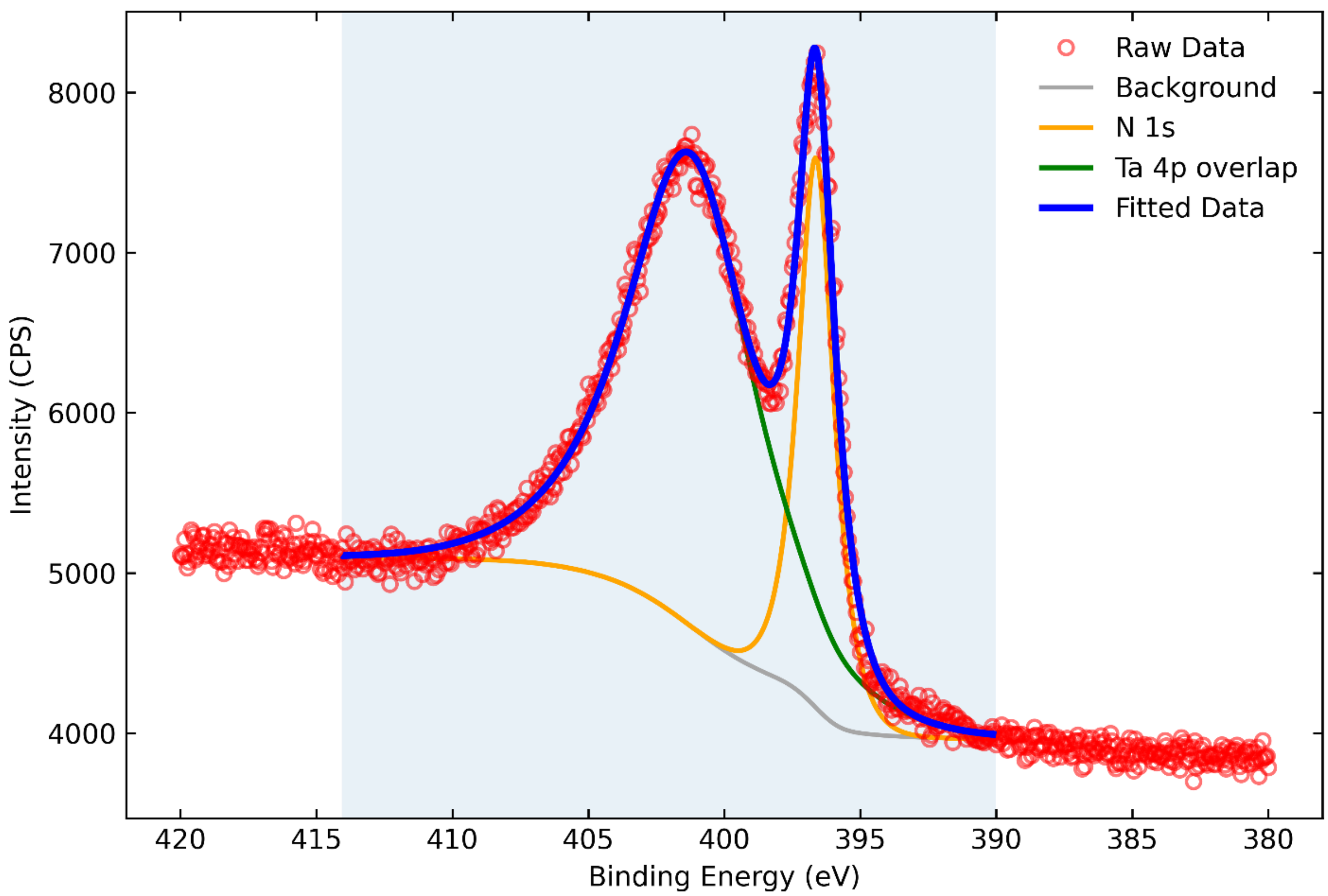

FIG. S13: N 1s XPS spectrum and fit for the 25 nm ALD TaN film after 60 s sputter time. The fitted peak areas were used to determine the N/Ta atomic ratio at this etch depth.